\documentclass[12pt]{article}
\usepackage{setspace}
\usepackage[normalem]{ulem}
\usepackage{fullpage,tweaklist}
\usepackage[letterpaper,top = 1in, bottom = .75in, left = 1in, right = 1in, nohead, includefoot, footskip= .25in]{geometry}
\usepackage[backref,colorlinks,citecolor=blue,bookmarks=true]{hyperref}
\usepackage{enumitem}
\usepackage{amsthm,amsmath, amssymb}
\usepackage{bbm}
\usepackage{pgfplots}
\pgfplotsset{compat=1.4}
\usetikzlibrary{patterns}
\usepackage{mdframed}
\renewcommand{\enumhook}{\setlength{\topsep}{0pt}%
  \setlength{\itemsep}{0pt}
  }
\renewcommand{\itemhook}
{
    \setlength{\topsep}{0pt}%
    \setlength{\itemsep}{0pt}

}

\definecolor{lightgray}{rgb}{0.95,0.95,0.95}
\mdfdefinestyle{boxed}{
innerleftmargin=5,innerrightmargin=5,
backgroundcolor=lightgray,
  topline=false,
  rightline=false,
  leftline=false,
  bottomline=false}

\newtheorem{definition}{Definition}
\newtheorem{example}{Example}
\newtheorem{lemma}{Lemma}
\newtheorem{theorem}{Theorem}
\newtheorem{corollary}{Corollary}

\newtheorem{proposition}{Proposition}
\theoremstyle{remark}

\newtheorem{observation}{Observation}

\newenvironment{prevproof}[2]{\noindent {\sc {Proof of {#1}~\ref{#2}:}}}{\hfill $\Box$\vskip \belowdisplayskip}

\newcommand{\cD}{{\cal D}}

\newcommand{\cR}{{\cal R}}

\newcommand{\cW}{{\cal W}}

\newcommand{\junk}[1]{}
\junk{

}
\newcommand{\notshow}[1]{}
\notshow{

}

\newcommand{\poly}{\text{poly}}

\DeclareMathOperator{\argmax}{argmax}
\DeclareMathOperator{\argmin}{argmin}

\definecolor{MyGray}{rgb}{0.9,0.9,0.9}

\begin{document}
\onehalfspacing

\title{{A Constructive Approach to Reduced-Form Auctions with Applications to Multi-Item Mechanism Design}}
\author {Yang Cai\thanks{Supported by NSERC Discovery RGPIN-2015-06127, FRQNT 2017-NC-198956 and NSF Awards CCF-0953960 (CAREER), CCF-1101491, and CCF-1617730. Work done in part while the author was a Research Fellow at the Simons Institute for the Theory of Computing.}\\
School of Computer Science, McGill \\
\tt{cai@cs.mcgill.ca}
\and
Constantinos Daskalakis\thanks{Supported by a Sloan Foundation Fellowship, a Microsoft Research Faculty Fellowship, and NSF Awards CCF-0953960 (CAREER),  CCF-1101491, and CCF-1617730. Work done in part while the author was a Research Fellow at the Simons Institute for the Theory of Computing.}\\
EECS, MIT \\
\tt{costis@mit.edu}
\and
S. Matthew Weinberg\thanks{{Research completed in part while the author was supported by a NSF Graduate Research Fellowship, and in part while the author was a Microsoft Research Fellow at the Simons Institute for the Theory of Computing.}}\\
Computer Science, Princeton\\
\tt{smweinberg@princeton.edu}
}
\maketitle

\begin{abstract}
We provide a constructive proof of Border's theorem~\cite{Border91,HartR11} and its generalization to reduced-form auctions with asymmetric bidders~\cite{Border07,ManelliV10,CheKM11}. Given a reduced form, we identify a subset of Border constraints that are necessary and sufficient to determine its feasibility. Importantly, the number of these constraints is linear in the total number of bidder types. 
In addition, we provide a characterization result showing that every feasible reduced form can be induced by an ex-post allocation rule that is a distribution over ironings of the same total ordering of the union of all bidders' types. 

We show how to leverage our results for single-item reduced forms to design auctions with heterogeneous items and asymmetric bidders with valuations that are additive over items. Appealing to our constructive Border's theorem, we obtain polynomial-time algorithms for computing the revenue-optimal auction. Appealing to our characterization of feasible reduced forms, we characterize feasible multi-item allocation rules.
\end{abstract}


\section{Introduction} \label{sec:intro}

Consider a mechanism design setting with $n$ bidders whose types lie in some finite set $T$ and one copy of a single indivisible item. In this setting, explicitly describing an ex-post allocation rule requires $|T|^n$ probability distributions. In particular, for every type profile one needs to specify the probability that the item is allocated to each bidder. In applications where a succinct characterization of optimal mechanisms is lacking\footnote{{Such applications include, for example, settings where bidders are risk-averse~\cite{MaskinR84}, budget-constrained~\cite{LaffontR96}, or have multi-dimensional preferences~\cite{RochetC98}.}} (c.f. revenue optimal auctions with independent quasi-linear and risk-neutral bidders~\cite{Myerson81}), it is desirable to take as a first step an optimization approach. However, optimizing over ex-post allocation rules is too expensive computationally and provides little structural insight into the optimal mechanism.

The above considerations motivate the study of {\emph{interim allocation rules}, also called \emph{reduced-form auctions}, or simply {\em reduced forms}. Formally, suppose that bidder $i$'s type is distributed according to some distribution $\cD_i$ over $T$, and that bidders' types are independent. The reduced form of an allocation rule is a collection  of functions ${\cal R}:=\{\pi_i: { T} \rightarrow [0,1]\}_{i\in[n]}$. For all bidders $i$ and types $\tau \in { T}$, $\pi_{i}(\tau)$ is the probability that the item is allocated to bidder $i$ conditioning on her report being $\tau$. The conditional probability is defined with respect to the reports of the other bidders, assumed to be drawn from the product distribution $\times_{j \neq i} \cD_j$, and any randomization used by the allocation rule itself.

A key question surrounding reduced forms is: \emph{under what conditions is a reduced form feasible?} More specifically, given a reduced form does there exist an ex-post allocation rule inducing it? This question was studied by Matthews~\cite{Matthews84} and Maskin and Riley~\cite{MaskinR84}, and Border~\cite{Border91} provided a collection of linear inequalities that are necessary and sufficient for the feasibility of a symmetric reduced form, namely when $\cD_i=\cD_j$ and $\pi_i(\cdot)=\pi_j(\cdot)$ for all bidders $i$ and $j$. In more recent work, Border~\cite{Border07}, Manelli and Vincent~\cite{ManelliV10}, and Che et al.~\cite{CheKM11} extend Border's conditions to the general (asymmetric) case. It follows from these works that a reduced form $\cR$ is feasible if and only if
\begin{align} \forall x_1,\ldots,x_n: \sum_{i} \sum_{\tau_i : \pi_i(\tau_i) \ge x_i}  \pi_i(\tau_i) \Pr_{t_i \sim \cD_i}[t_i=\tau_i] \le 1-\prod_{i}\left(1-\Pr_{t_i \sim \cD_i}[\pi_i(t_i) \ge x_i]\right).\label{eq:improved feasibility condition non symmetric}\end{align}
Intuitively, the left hand side of~\eqref{eq:improved feasibility condition non symmetric} represents the probability that the item is allocated to some bidder $i$ whose realized type $\tau_i$ satisfies $\pi_i(\tau_i) \ge x_i$, as computed by the reduced form $\cR$. The right hand side represents the probability that there exists some bidder $i$ whose realized type $\tau_i$ satisfies $\pi_i(\tau_i) \ge x_i$. Clearly, if $\cR$ is feasible, it outght to satisfy~\eqref{eq:improved feasibility condition non symmetric} for any choice of thresholds $x_1,\ldots,x_n$. What is less clear is that if the inequality is satisfied for all thresholds, then $\cR$ is feasible (but indeed this is the case).

From an optimization standpoint, one drawback of the afore-described conditions is that there are about $|T|^n$ inequalities that one needs to verify. Our first main result is that in fact it suffices to only check a subset of $n |T|$ constraints to verify the feasibility of a given reduced form.
\begin{theorem}\label{thm: independent} A reduced form $\cR$ is feasible if and only if
\begin{align} \forall x: \sum_{i} \sum_{\tau_i \in S^{(i)}_x}  \pi_i(\tau_i) \Pr_{t_i \sim \cD_i}[t_i=\tau_i] \le 1-\prod_{i}\left(1-\Pr_{t_i \sim \cD_i}\left[t_i \in S^{(i)}_x\right]\right),\label{eq:our virtual feasibility condition}
\end{align}
where $S^{(i)}_x = \left\{\tau_i \in T~|~ {\pi}_i(\tau_i) \cdot \Pr_{t_i \sim \cD_i}[\pi_i(\tau_i) \ge \pi_i(t_i)]>x\right\}$.
In particular, one can test the feasibility of a reduced form or obtain a hyperplane separating $\cR$ from the set of feasible reduced forms in {time {$O(|T| n \cdot \log (|T|n))$}}.
\end{theorem}
Note that the collection of inequalities~\eqref{eq:our virtual feasibility condition} are a subset of  inequalities~\eqref{eq:improved feasibility condition non symmetric}. In particular, we have only kept $n|T|$ out of about $|T|^n$ inequalities. One way to interpret our theorem is that it coordinates which combinations of thresholds $x_1,\ldots,x_n$ it suffices to check simultaneously in~\eqref{eq:improved feasibility condition non symmetric}. A priori, the simplest approach that could conceivably work would be to set all the $x_i$'s equal, but we show that this does not suffice; see Example Three in Section~\ref{sec:examples}. Instead of comparing different bidders' interim probabilities of allocation at face value, our theorem describes a way to shade these probabilities depending on each bidder's type distribution. The resulting ``shaded interim probabilities'' {\em can} be compared at face value. {We provide several examples showing how Theorem~\ref{thm: independent} can be used to turn the task of verifying the feasibility of interim allocation rules analytically tractable in Section~\ref{sec:examples}.}

In addition to determining the feasibility of reduced forms, it is also important to understand the structure of ex-post allocation rules that induce them. To this end, Manelli and Vincent~\cite{ManelliV10} provide an interesting characterization result, using the notion of a {\em hierarchical allocation rule}~\cite{Border91}. A hierarchical allocation rule maintains a weak total ordering $\succeq$ over the elements of $[n] \times T \cup \{(0,\bot=\tau_0)\}$. On input $(\tau_1,\ldots,\tau_n)$, the allocation rule computes the subset of indices {$\cW =\{i~|~(i,\tau_i) \succeq (j,\tau_j), \forall j\}$, then selects a uniformly random index $i$ in $\cW$. If $i > 0$, the item is allocated to bidder $i$. If $i = 0$, the item is not allocated.} Manelli and Vincent show that, if a reduced form $\cR$ is feasible, then there exists a distribution over hierarchical allocation rules inducing it. Moreover, each hierarchical allocation rule in the support of the distribution uses a weak total ordering $\succeq$ satisfying $\pi_i(\tau_i') \ge \pi_i(\tau_i) \implies (i,\tau_i') \succeq (i,\tau_i)$, i.e. ``stronger types'' of bidder $i$ are ranked higher than ``weaker types'' of bidder $i$ in every hierarchical allocation rule in the support. We strengthen this characterization as follows.
\begin{theorem} \label{thm:non-iidhierarchical}
If a reduced form $\cR$ is feasible then there exists a strict total ordering $\succ$ over the elements of $[n] \times T \cup {\{(0,\bot)\}}$ such that $\cR$ can be induced by a distribution over hierarchical allocation rules, each using a weak ordering that irons $\succ$.\footnote{We say that a weak ordering $\succeq$ over some set $S$ ``irons'' a strict ordering $\succ'$ over the same set $S$ iff, for all $i,j \in S$, $i \succ' j \implies i \succeq j$.} 
\end{theorem}

In addition to the \emph{per-bidder} notion of ``strength'' of types guaranteed by Manelli and Vincent's characterization, Theorem~\ref{thm:non-iidhierarchical} guarantees the existence of a \emph{global} notion of strength of types in the sense that for {every} profile $(\tau_1,\ldots,\tau_n)$ with $\tau_{i_1} \succ \tau_{i_2} \succ \ldots \succ\tau_{i_n}$, the ex-post allocation probabilities $p_1,\ldots,p_n$ of the item to the bidders satisfy $p_{i_1} \ge p_{i_2} \ge \ldots \ge p_{i_n}$.

\subsection{Multi-Dimensional Mechanism Design}

Designing revenue-optimal auctions in multi-item settings has been a challenging application domain in mechanism design. A characterization theorem \`{a} la Myerson~\cite{Myerson81} is unknown, and it is well-understood that optimal multi-item mechanisms exhibit much richer structure compared to optimal single-item auctions, involving bundling and randomization even in the case of a single additive bidder;\footnote{A bidder is additive if her valuation for a set of items is equal to the sum of her values for each item in that set.} see discussion in Section~\ref{sec:related}. In light of this, it is valuable to develop optimization tools to compute optimal multi-item mechanisms. 

Towards this end we adopt a linear programming approach. It is easy to write a linear program optimizing expected revenue over ex-post allocation and price rules of feasible, Bayesian incentive compatible mechanisms. However, this approach has two drawbacks. First, describing ex-post allocation and price rules requires $(m+1)n|T|^n$ numbers {(for each of $|T|^n$ type profiles, and each of $n$ bidders, one must list an allocation probability for each of $m$ items along with a price paid)}. The exponential dependence on $n$ makes this approach computationally intractable. And, even if we could solve this linear program, it would be hard to discern useful structural insights from a ``laundry list'' of allocation probabilities and prices for every type profile.

The afore-described difficulties motivate a linear programming formulation with respect to the interim description of a mechanism instead. To write such a linear program, we need to identify linear constraints guaranteeing that an interim allocation/price rule pair is both feasible and Bayesian Incentive Compatible. When bidders are additive, Bayesian Incentive Compatibility is easy to express directly in terms of the interim allocation and price rule with a short list of $n |T|^2$ linear constraints. What is not clear is how to concisely express the feasibility of a multi-item interim allocation rule.

Our key observation is the following: a multi-item interim allocation rule is feasible if and only if the single-item interim rules that it projects onto each item are all feasible. Therefore, it suffices to invoke our single-item results above to resolve this problem. {To be absolutely clear, even when the bidders are additive, it is folklore knowledge and well-understood that in the revenue-optimal auction \textbf{the interim probability that a bidder receives some item must in principle depend on her values for the other items.} It is exactly this property that makes multi-dimensional mechanism design notoriously difficult and not just a product of tractable single-item problems, and we are not claiming otherwise. However, the very specific subproblem of \textbf{determining whether a multi-item interim allocation rule is feasible can be solved separately across items}}. Making use of Theorem~\ref{thm: independent} as a subroutine inside a linear program solver, we obtain the following computational result:

\begin{theorem}\label{thm:computation main} There is a polynomial-time algorithm that finds a revenue-optimal, BIC mechanism in multi-item settings with additive bidders. The algorithm takes as input the type distributions $\cD_1,\ldots,\cD_n$ of the bidders, and outputs a concise description of an optimal mechanism in time polynomial in the number of bidders $n$, the number of items $m$ and the size of the type-space, $|T|$. The bidders are assumed independent, but each $\cD_i$ may be an arbitrarily correlated distribution over item values.
\end{theorem}

Besides Theorem~\ref{thm:computation main}, our key observation stated above, combined with Theorem~\ref{thm:non-iidhierarchical}, directly implies a characterization of feasible multi-item interim allocation rules, as follows.

\bigskip \noindent 
\colorbox{MyGray}{
\begin{minipage}{\textwidth} {{\bf Characterization of Feasible Multi-Item Interim Allocation Rules}

Every feasible multi-item interim allocation rule {can be implemented as follows}:
\begin{itemize}
\item Every item is allocated independently of the other items.
\item The allocation rule of every item $\ell$ maintains:
\begin{itemize}
\item a strict ordering $\succ_\ell$ over the elements of $[n]\times T \cup {\{(0,\bot{=\tau_0})\}}$; and
\item a distribution over ironings of $\succ_\ell$.
\end{itemize}
\item Each item is then allocated as follows. First, a random ironing $\succeq_\ell'$ is sampled. On an input of reported types $(\tau_1,\ldots,\tau_n)$, the allocation rule computes the subset of indices {$\cW_\ell =\{i~|~(i,\tau_i) \succeq_\ell' (j,\tau_j), \forall j\}$, then selects a uniformly random index $i_\ell$ in $\cW_\ell$. If $i_\ell > 0$, item $\ell$ is allocated to bidder $i_\ell$. If $i_\ell = 0$, item $\ell$ is not allocated.}
\end{itemize}}
\end{minipage}} 
\smallskip

Recall that the set $T$ in the characterization above is the set of types a bidder may have. In particular, each element  $\tau \in T$ determines the values of a bidder of type $\tau$ for each bundle of items. Hence, the content of the first bullet is that \textbf{while the ex-post allocation rule for item $\ell$ indeed must depend on bidders' values for items $\neq \ell$, it need not depend on how items $\neq \ell$ are themselves allocated}.


\subsection{Related Work} \label{sec:related}
\paragraph{Reduced Forms.} A necessary and sufficient condition for the feasibility of a bidder-symmetric reduced form was provided by Border~\cite{Border91}, building on prior work by Maskin and Riley~\cite{MaskinR84} and Matthews~\cite{Matthews84}. A simpler proof of Border's theorem and alternative criteria for feasibility were also provided by Hart and Reny~\cite{HartR11}. {For all these works, the necessary and sufficient conditions take the form of $|T|$ linear inequalities}. Border's conditions for the symmetric setting were generalized to the asymmetric setting by Border~\cite{Border07}, Manelli and Vincent~\cite{ManelliV10}, and Che et al.~\cite{CheKM11}. {For these works, the necessary and sufficient conditions take the form of $|T|^n$ linear inequalities.\footnote{{To be more precise, Border's~\cite{Border07} conditions took the form of $2^{|T|n}$ linear inequalities, and Che et al.~\cite{CheKM11} identified a necessary and sufficient subset of $|T|^n$ linear inequalities.}}} {In comparison, Theorem~\ref{thm: independent} shows that $|T|n$ linear inequalities suffice.}

Let us review in more detail some of the most related works on reduced forms. Manelli and Vincent characterize the extreme points of the space of feasible, monotone reduced forms as monotone hierarchical allocation rules in both the bidder-symmeteric and asymmetric case when bidders have continuous type spaces. This implies that every monotone reduced form has an ex-post allocation rule inducing it that is also ex-post monotone. In this work, we offer alternative proofs of these results (Theorems~\ref{thm:iidhierarchical} and~\ref{thm: independentcorners}) when type spaces are finite. {Owing to the simpler nature of finite versus infinite dimensional geometry (in particular that counting arguments over finite sets are considerably simpler than counting arguments over infinite sets), our proofs more clearly isolate the key insights and are considerably shorter. Therefore, we include these proofs both for the sake of completeness and intuition.} {As detailed previously, in comparison to these characterization results, Theorem~\ref{thm:non-iidhierarchical} provides a stronger characterization of reduced forms as implementable via randomizations over hierarchical allocation rules that iron the same \emph{global} total ordering of all bidders' types.}

Che et al. provide a clean network-flow interpretation of Border's theorem for asymmetric bidders, and show how to also accommodate bidders' capacity constraints in multi-unit generalizations {(for example, that the set $G$ of bidders must never receive more than $C(G)$ units or less than $L(G)$ units on any profiles). Their necessary and sufficient conditions remain in the form of $|T|^n$ linear inequalities (same as for a single item with asymmetric bidders) despite the significant increase in generality where their results apply}. Independently from our work, Alaei et al.~\cite{AlaeiFHHM12} also provide a computationally efficient algorithm to determine the feasibility of reduced forms via a ``token-passing game.'' Their work shows in fact that there exists a collection of roughly $(n|T|)^2$ inequalities that define the space of feasible reduced forms. Their characterization is what is called an ``extended formulation'' - they introduce an additional $(n|T|)^2$ variables and their inequalities are not of the form~\eqref{eq:improved feasibility condition non symmetric}. The results of Alaei et al. further extend to settings where the allocation is subject to matroid constraints. We address neither capacity constraints nor matroid constraints in this work. {One high-level distinction between the main contributions of these works and our Theorems~\ref{thm: independent} and~\ref{thm:non-iidhierarchical} is that their results extend Border's theorem to more general settings, whereas our work provides deeper insight into the core single-item setting (and by extension, as observed earlier, the multi-item setting where each item can be feasibly allocated to any bidder, regardless of other items she is allocated)}. } 

{Finally, several recent papers have provided polynomial-time algorithms for determining whether a reduced form is \emph{approximately} feasible in multi-item settings with more complex allocation constraints~\cite{CaiDW12b,CaiDW13a,CaiDW13b}. On the other hand, Gopalan et al. show essentially that approximation is the best one can hope for: their work identifies a formal barrier to the existence of exact and ``computationally useful'' Border-like theorems beyond single-item settings~\cite{GopalanNR15}.}

\paragraph{Multi-Item Auctions.} Prior work on multi-dimensional mechanism design is extensive (see e.g. survey~\cite{ManelliV07}), driven by the scarcity of settings where the optimal mechanism has a clean allocation rule (such as Myerson's revenue-optimal auction for single-dimensional settings~\cite{Myerson81}, or the welfare-optimal VCG auction in quite general settings~\cite{Vickrey61, Clarke71, Groves73}). Indeed, numerous formal barriers have been identified to the existence of clean, revenue-optimal multi-item mechanisms, such as the necessity of randomization~\cite{RochetC98,Thanassoulis04, Pavlov11}, large menu complexity~\cite{BriestCKW10,HartN13, DaskalakisDT13,DaskalakisDT16}, {large description complexity~\cite{DaskalakisDT14}}, and revenue non-monotonicity~\cite{HartR12,RubinsteinW15}. Daskalakis et al.~\cite{DaskalakisDT16} have recently provided a characterization of single-bidder revenue-optimal mechanisms using optimal transport theory. Older work of Rochet and Chon\'{e}~\cite{RochetC98} had provided a characterization of optimal single-bidder mechanisms in the related setting where there is no bound on the number of units per item but the seller has a strictly convex production cost for generating more units. Finally, the problem has recently entered the Theory of Computation, where the emphasis has mostly been in deriving computationally efficient algorithms for computing optimal mechanisms. A number of results have emerged obtaining constant-factor approximations in polynomial time~\cite{CaiD11, Alaei11,BhattacharyaGGM10,ChawlaHMS10, ChawlaMS15, KleinbergW12, CaiH13, BabaioffILW14, Yao15, ChawlaM16, CaiZ17}. 

{In comparison to these works, ours is the first to provide a poly-time algorithm and corresponding characterization of revenue-optimal multi-item auctions without any distributional assumptions (such as a hazard rate condition {or item-value independence}). Indeed, following the announcement of portions of this work~\cite{CaiDW12a}, several works (including some by the authors) provided computationally efficient algorithms to find approximately-optimal mechanisms in increasingly general multi-item settings~\cite{CaiDW12b,CaiDW13a,CaiDW13b, BhalgatGM13, DaskalakisDW15, CaiDW16}.} {In comparison to these works, the present paper remains unique in containing a computationally efficient algorithm to find the exact optimal mechanism in a multi-item setting without any approximation error.}

\subsection{Roadmap of Remaining Sections} Section~\ref{sec:notation} below makes clear the notation we use and formal questions we study with respect to reduced forms. Section~\ref{sec:iid} studies bidder-symmetric reduced forms as a warm-up for Section~\ref{sec:independent}, which studies asymmetric reduced forms. Section~\ref{sec:characterization} provides our results on multi-item auctions. {Appendix~\ref{app:proofs} contains some omitted proofs.}

\section{Preliminaries and notation}\label{sec:notation}
Throughout the paper, we denote the number of bidders by $n$. We also use $T$ to denote the possible types of a bidder. {In order to obtain computationally meaningful results, we assume that $T$ is finite and use $c$ as shorthand for $|T|$, but make no other assumptions on $T$.} In particular, it is not necessary to assume that $T$ is a subset of $\mathbb{R}$.

We use $\tau$ to denote the \emph{type} of a bidder, without emphasizing whether it is a vector or a scalar (or otherwise).  The elements of $T^n$ are called {\em type profiles}, and specify a type for every bidder. We assume type profiles are sampled from {a distribution ${\cal D}=\prod_{i=1}^{n}{\cal D}_{i}$ over $T^n$}, where ${\cal D}_i$ the marginal of this distribution on bidder $i$'s type, and use ${\cal D}_{-i}$ to denote the marginal distribution over the types of all bidders, except bidder $i$.  We use $t_i$ for the random variable representing the type of bidder~$i$. So when we write $\Pr[t_i = \tau]$, we mean the probability that bidder $i$'s type is $\tau$. If bidders are i.i.d., because $\Pr[t_i = \tau]$ is the same for all $i$, we will just write $\Pr[\tau]$.

The {\em reduced form} $\cR$ of an allocation rule specifies {a vector function $\pi(\cdot)$,} specifying values $\pi_{i}(\tau)$, for all bidders $i$ and types $\tau \in T$. $\pi_{i}(\tau)$ is the probability that bidder $i$ receives the item when reporting type $\tau$, where the probability is over the randomness of all other bidders' types and the internal randomness of the allocation rule, assuming that the other bidders report their true types. We may think of $\cR$ as a vector in {$[0,1]^{nc}$, by simply listing $\pi_i(\tau)$ for all $i, \tau$, and will sometimes write $\vec{\pi}$ to emphasize this view. 

In Section~\ref{sec:iid}, we consider settings where the bidders are i.i.d., i.e. $\cD_i = \cD_j$ for all $i, j$, and the reduced forms are bidder-symmetric, which satisfy $\pi_{i}(\tau) = \pi_{j}(\tau)$, for all $i, j, \tau $. In such cases, we will drop the subscript $i$, writing just $\pi(\tau)$, for the probability that a bidder of type $\tau \in T$ receives the item, over the randomness of the allocation rule and the types of the other bidders, {assuming that the other bidders report their true types.} Given a reduced form $\cR$, we will be interested in whether it is ``feasible.''  By this, we mean ``does there exist an ex-post allocation rule that never over-allocates the item whose reduced form is $\cR$?'' If the answer to this question is ``yes'' then we will also say that the ex-post allocation rule whose reduced form is $\cR$, ``induces $\cR$'' or ``implements $\cR$.'' {Note that there is some subtlety in defining feasibility of a reduced form if $\Pr[t_i = \tau] = 0$ for some $i \in [n], \tau \in T$. There are a couple reasonable choices that are qualitatively the same - we choose to define a reduced form to be feasible only if $\Pr[t_i = \tau] = 0 \Rightarrow \pi_i(\tau) = 0$.}

We also note that the running times of the algorithms obtained in Sections~\ref{sec:iid} and~\ref{sec:independent} are quoted without accounting for the{ \emph{bit complexity} of numbers involved. The bit complexity of a rational number $x$ is the number of bits $b$ required so that $x$ can be expressed as the ratio of two binary numbers with $b$ bits of precision. If $b$ upper bounds the bit complexity of $\Pr[t_i = \tau]$ for all $i, \tau$, and all coordinates of the input reduced form $\vec{\pi}$,} then it suffices to multiply all quoted running times by a factor that is polynomial in $b$.

\paragraph{Order Notation.} Throughout the text we use the $O(\cdot)$ notation. Let $f(x)$, $g(x)$ be two positive functions defined on some infinite subset of $\mathbb{R}_+$. Then we write $f(x) = O(g(x))$ iff there exist some positive reals $\alpha$ and $x_0$ such that $f(x) \le \alpha g(x)$, for all $x>x_0$. {We also write $f(x) = \poly(x)$ iff there exist positive reals $\alpha$ and $x_0$ such that $f(x) \leq x^\alpha$, for all $x > x_0$. }

\vspace{.1in}
\noindent Finally, we provide some brief geometric preliminaries. 

\begin{definition}(\textbf{Corner}) Let $P$ be a closed, convex subset of Euclidean space defined as the intersection of finitely many halfspaces. Namely, $P = \cap_{i \in {\cal I}} \{ \vec{x} | \vec{a}_i \cdot  \vec{x} \le b_i \},$ for some finite index set ${\cal I}$. We say that $\vec{x}^*$ is a \emph{corner} of $P$ if $\vec{x}^* \in P$, and the set of equations $\{\vec{a}_i \cdot \vec{x} = b_i\}_{i \in S}$ has as a unique solution the point $\vec{x}^*$, where $S = \{i\in \mathcal{I} | \vec{a}_i \cdot \vec{x}^* = b_i\}$. 

\end{definition}

\begin{definition}(\textbf{Separation Oracle}) Let $P$ be a closed, convex subset of Euclidean space. Then a \emph{Separation Oracle} for $P$ is an algorithm that takes as input a point $\vec{x}$ and outputs ``\textsc{Yes}'' if $\vec{x} \in P$, or a hyperplane $(\vec{w},c)$ such that $\vec{y} \cdot \vec{w} \leq c$ for all $\vec{y} \in P$, but $\vec{x} \cdot \vec{w} > c$. Note that because $P$ is closed and convex, such a hyperplane always exists whenever $\vec{x} \notin P$. 

{A separation oracle is \emph{poly-time} if on inputs of bit complexity $b$, it terminates in time $\poly(b, x)$, where $x$ is the maximum bit complexity of any coordinate in any halfspace defining $P$. }

\end{definition}

\noindent We will also make use of the following theorem, reworded from~\cite{Khachiyan79,GrotschelLS81, KarpP82}.

\begin{theorem}\label{thm:GLS}(\cite{Khachiyan79,GrotschelLS81,KarpP82}) Let $P$ be a $d$-dimensional closed, convex subset of $\mathbb{R}^d$ defined as the intersection of finitely many halfspaces, and $SO$ be a {poly-time} separation oracle for $P$. Then it is possible to do the following:
\begin{itemize}
\item Find an element in $\argmax_{\vec{x} \in P} \{\vec{c} \cdot \vec{x}\}$ for any $\vec{c} \in \mathbb{Q}^d$ (i.e. solve linear programs) in time polynomial in $d$, {and $b$, where $b$ upper bounds the bit complexity of all coordinates of the vector $\vec{c}$, and all coordinates of the halfspaces defining $P$.} 

\item Decompose any $\vec{x} \in P$ into a convex combination of at most $d+1$ corners of $P$ in time polynomial in $d$, {and $b$, where $b$ upper bounds the bit complexity of all coordinates of the vector $\vec{x}$, and all coordinates of the halfspaces defining $P$.} 
\end{itemize}
\end{theorem}

\section{Warm-Up: Bidder-Symmetric Reduced Forms} \label{sec:iid}
{This section serves as a warm-up for our main results by viewing bidder-symmetric reduced forms through a computational lens. Some of the key ideas for our main results in Section~\ref{sec:independent} can be more cleanly illustrated for symmetric bidders below.}
 
Let us begin by reviewing Border's theorem~\cite{Border91}, which specializes~\eqref{eq:improved feasibility condition non symmetric} to the case of i.i.d. bidders and bidder-symmetric reduced forms. A bidder-symmetric reduced form is feasible if and only if:
\begin{align}\forall x:~~ n \cdot \sum_{\tau: \pi(\tau) \geq x} \pi(\tau) \Pr[\tau] \le 1-\left(1-\Pr_{t \sim \cD_1}[\pi(t) \geq x]\right)^n.\label{eq:border's condition}\end{align}

{The semantic meaning of the above equations are the same as those in Equation~\eqref{eq:improved feasibility condition non symmetric}: the left-hand side denotes the probability that some bidder whose type $\tau$ satisfies $\pi(\tau) \geq x$ receives the item, as promised by the reduced form, and the right-hand side denotes the probability that some bidder $i$ has $\pi(t_i) \geq x$ as computed by the probability distribution. Note that there are drastically fewer inequalities to check of form~\eqref{eq:border's condition}: only $|T|$ instead of $|T|^n$. This is essentially because if there exist thresholds $x_1,\ldots,x_n$ for which an equation of form~\eqref{eq:improved feasibility condition non symmetric} is violated, there is also a single $x$ such that Equation~\eqref{eq:border's condition} is violated at $x$. As there are only $|T|$ inequalities to check, Equation~\eqref{eq:border's condition} directly implies Corollary~\ref{cor:algiid} below: one can determine the feasibility of a reduced form in{ time $O(c(\log c + \log n))$} via a routine computation. A proof is included in Appendix~\ref{app:proofs}.}
\begin{corollary}[of~\cite{Border91}]\label{cor:algiid}
The feasibility of a given bidder-symmetric reduced form can be determined in time $O(c(\log c + \log n))$. If it is infeasible, a violated inequality of form~\eqref{eq:border's condition} can be determined in the same time.
\end{corollary}

Now that it is easy to determine the feasibility of a reduced form, we wish to understand how to find an ex-post allocation rule inducing a given feasible reduced form in poly-time. To this end, let us formally define hierarchical allocation rules, again specialized to the bidder-symmetric case.

\begin{definition}\label{def:hierarchy} (\cite{Border91})
A \textbf{hierarchical allocation rule} consists of a weak total ordering $\succeq$ over $T \cup \{\bot\}$. On reports $(\tau_1,\ldots,\tau_n)$, the allocation rule computes the subset of indices $\cW =\{i\geq 1~|~\tau_i \succeq \bot~\text{and}~\tau_i \succeq \tau_j, \forall j\}$, then selects a uniformly random bidder $i$ in $\cW$, if $\cW$ is non-empty. If $\cW$ is empty, the item is unallocated.

We say that a hierarchical allocation rule induced by $\succeq$ is \emph{well-ordered} with respect to a reduced form $\cR$ if $\pi(\tau) \geq \pi(\tau') \Rightarrow \tau \succeq \tau'$. \end{definition}
For every feasible reduced form $\cR$, Theorem~\ref{thm:iidhierarchical} below characterizes the corners of a convex region containing it - which is intimately connected to ex-post allocation rules inducing $\cR$.

\begin{theorem}(implied by~\cite{ManelliV10})\label{thm:iidhierarchical}
Every feasible reduced form $\cR$ lies inside a $c$-dimensional polytope $P$ whose corners are all {reduced forms of} hierarchical allocation rules that are well-ordered w.r.t. $\cR$. Furthermore, there is a distribution over at most $c+1$ hierarchical allocation rules, all well-ordered w.r.t. $\cR$, that induces $\cR$.
\end{theorem}

\begin{proof} 
For ease of notation, first relabel all types in $T$ so that ${\pi}(\tau_1) \geq {\pi}(\tau_2) \ldots \geq {\pi}(\tau_{c})$. Let $S = \{i | {\pi}(\tau_i) = {\pi}(\tau_{i+1})\}$ (for notational convenience, denote by $\pi(\tau_{c+1}) = 0$). {Consider the convex polytope $P\subseteq [0,1]^{c}$ specified by the following constraints.} 
\begin{align}
\tilde{\pi}(\tau_i) &= \tilde{\pi}(\tau_{i+1}) &\forall i \in S; \\
\tilde{\pi}(\tau_i) &\geq \tilde{\pi}(\tau_{i+1}) &\forall i \in [c]-S; \\
\sum_{j \leq i} n\cdot \Pr[t_j]\tilde{\pi}(\tau_j) &\leq 1 - \left(1-\sum_{j\leq i}\Pr[t_j]\right)^n &\forall i \in [c];
\end{align}
where for notational convenience we denote $\tilde{\pi}(\tau_{c+1}) = 0$ ({so in particular $\tilde{\pi}(\tau_1), \cdots, \tilde{\pi}(\tau_{c})$ are the free variables and $\tilde{\pi}(\tau_{c+1})$ is not,} and the afore-desribed polytope is a subset of $[0,1]^c$). {By~\eqref{eq:border's condition}, $\cR$ is feasible if and only if $\pi\in P$.} Consider the corners of this polytope. As there are $c$ variables, every corner must satisfy at least $c$ of the above inequalities with equality. Refer to the constraint $\tilde{\pi}(\tau_i) \geq \tilde{\pi}(\tau_{i+1})$ or $\tilde{\pi}(\tau_i) = \tilde{\pi}(\tau_{i+1})$ (whichever is included in the definition of $P$) as the $i^{th}$ monotonicity constraint, and the constraint $\sum_{j \leq i} n\cdot \Pr[t_j]\tilde{\pi}(\tau_j) \leq 1 - \left(1-\sum_{j\leq i}\Pr[t_j]\right)^n$ as the $i^{th}$ Border constraint. 

We first show that no feasible reduced form satisfies both the $i^{th}$ monotonicity constraint \emph{and} the $i^{th}$ Border constraint with equality. Consider then a feasible reduced form $\tilde{\pi}$, and an ex-post allocation rule $\tilde{M}$ inducing $\tilde{\pi}$. Recall that if the $i^{th}$ Border constraint is tight, then the probability that a type in $\{\tau_1,\ldots,\tau_i\}$ receives the item is exactly the probability that such a type is reported to $\tilde{M}$. {Therefore, whenever one or more types from this set are reported to $\tilde{M}$, a bidder with a type from this set necessarily wins the item. In particular, t}his means that we must have {$\tilde{\pi}(\tau_i) \geq (1-\sum_{j \leq i}\Pr[t_j])^{n-1}$}, as $\tau_i$ must certainly win whenever all other reported types have index strictly larger than $i$. In fact, we must have $\tilde{\pi}(\tau_i) > (1-\sum_{j \leq i}\Pr[t_j])^{n-1}$, as $\tau_i$ must also win with non-zero probability in the disjoint event that there is at least one other reported type equal to $\tau_i$, and the remaining types have indicies strictly larger than $i$ (because $\cR$ is bidder-symmetric). Similarly, we necessarily have {$\tilde{\pi}(\tau_{i+1}) \leq (1-\sum_{j \leq i}\Pr[t_j])^{n-1}$}, as $\tau_{i+1}$ can \emph{only} win in the event that all other types have index strictly larger than $i$. Therefore, if the $i^{th}$ Border constraint is tight, the $i^{th}$ monotonicity constraint is not. 

Now let us consider a corner $\tilde{\pi}$ of $P$. Since $P$ lies in $\mathbb{R}^c$, there must be at least $c$ constraints that $\tilde{\pi}$ satisfies with equality. As there are {$2c$ constraints in total}, and no feasible reduced form can satisfy both the $i^{th}$ Border constraint and the $i^{th}$ monotonicity constraint, we see that every corner must satisfy either the $i^{th}$ Border constraint or the $i^{th}$ monotonicity constraint with equality. Such a reduced form corresponds to a hierarchical allocation rule with $\tau_i \succeq \tau_{j}$ for all $i< j$, and $\tau_j \succeq \tau_i$ for some $i < j$ if and only if the $k^{th}$ monotonicity constraints are tight for all $k \in \{i,\ldots, j-1\}$. Additionally, if $\tilde{\pi}(\tau_i) = 0$, 
{then $\bot \succeq \tau_i$ and $\tau_i \not\succeq \bot$. If $\tilde{\pi}(\tau_i) > 0$, then $\tau_i \succeq \bot$ and $\bot \not \succeq \tau_i$.} It is easy to see that this hierarchical allocation rule has a feasible reduced form that satisfies the desired equalities (hence it equals $\tilde{\pi}$) and is well-ordered w.r.t. $\cR$.

{By Carath\'eodory's theorem, we can write any feasible reduced form $\pi$ as ${\pi}=\sum_{j=1}^{c+1} w_{j}\tilde{\pi}_{j}$, where $\sum_{j=1}^{c+1} w_{j}=1$, and for all $j$, $w_{j}\geq 0$ and $\tilde{\pi}_{j}$ is a corner of $P$.} The last {step of the proof} is an immediate consequence of the following observation.

\begin{observation}\label{obs:distribution}
If a reduced form $\cR$ can be written as ${\pi}  = \sum_j w_j \tilde{\pi}_j$, where each $w_j \geq 0$ and $\sum_j w_j = 1$, and each $\tilde{\pi}_j$ is induced by the ex-post allocation rule $M_j$, then $\cR$ is induced by the ex-post allocation rule $\sum_j w_j M_j$ (sample $j$ with probability $w_j$, then use $M_j$).
\end{observation}

\vspace{-1.8em}
\end{proof}

\begin{corollary}\label{cor:grand theorem symmetric}
Given a bidder-symmetric reduced form $\cR$ we can determine if it is feasible, or find a hyperplane separating it from the set of feasible bidder-symmetric reduced forms in time {$O\left(c (\log c+ \log n)\right)$}. If $\cR$ is feasible, we provide a succinct description of an ex-post allocation rule inducing it, in time polynomial in $c$ and $\log n$. {In particular, the allocation rule is a distribution of at most $c+1$ hierarchical allocation rules, all well-ordered w.r.t. $\cR$}.
\end{corollary}

\begin{proof} The first sentence immediately follows from Corollary~\ref{cor:algiid}, and the observation that any violated Border inequality is exactly a hyperplane separating $\cR$ from the space of feasible bidder-symmetric reduced forms.

We now need to describe how to computationally efficiently find an ex-post allocation rule implementing a reduced form ${\pi}$ that is feasible. By Theorem~\ref{thm:iidhierarchical}, any feasible ${\cR}$ lies inside a $c$-dimensional polytope $P$ whose corners are all {reduced forms of} hierarchical allocation rules that are well-ordered w.r.t. $\cR$. {We now observe that we have defined a separation oracle for $P$ in the first paragraph that runs in time $O(c (\log c + \log n))$. So Theorem~\ref{thm:GLS} implies that we may decompose $\cR$ into a convex combination of corners of $P$ in time polynomial in $c$ and $c (\log c+\log n)$ (resulting in a runtime polynomial in both $c$ and $\log n$). Observation~\ref{obs:distribution} completes the proof.}

Notice in particular that in the proof of Theorem~\ref{thm:iidhierarchical}, we described an easy procedure to define a hierarchical allocation rule that implements any corner $\tilde{\pi}$ of $P$ in terms of the inequalities of the polytope $P$ that are tight at $\tilde{\pi}$.
\end{proof}

\section{Asymmetric Reduced Forms} \label{sec:independent}
{Let's begin this section by recapping the major components leading to Corollary~\ref{cor:grand theorem symmetric} for symmetric reduced forms:
\begin{enumerate}
\item First, we need a computationally-efficient algorithm that takes as input a prospective reduced form and finds a violated Border constraint, if it exists (and otherwise claims that all Border constraints are satisfied). For symmetric reduced forms, Border's Theorem~\cite{Border91} gives us this for free as there are only $|T|$ constraints to check. For asymmetric reduced forms, it will take exponential time to check all $|T|^n$ constraints of form~\eqref{eq:improved feasibility condition non symmetric}, so we show in Section~\ref{sec:shaded} that in fact it suffices to check only $n|T|$ constraints by properly shading the interim probabilities. 
\item Next, to implement feasible reduced forms, we need to understand the corners of a convex region containing all feasible reduced forms. Work of Manelli and Vincent accomplishes this for the symmetric case (Theorem~\ref{thm:iidhierarchical}) and the asymmetric case (Theorem~\ref{thm: independentcorners}) for continuous type spaces. Again, we include a proof for the asymmetric case in Section~\ref{sec:asymmetrichierarchical} for finite type spaces that follows the same intuitive approach as our proof of Theorem~\ref{thm:iidhierarchical}. 
\item Finally, we want to gain more insights into the structure of the space of feasible reduced forms than only understanding the corners of the feasible region. \notshow{understand just how much structure we can claim on the space of feasible reduced forms.} In the symmetric case, it is unclear what one might hope for beyond Theorem~\ref{thm:iidhierarchical}. In the asymmetric case, however, Theorem~\ref{thm: independentcorners} doesn't tell the whole story. More specifically, Theorem~\ref{thm: independentcorners} proves that every feasible reduced form can be induced by a distribution over hierarchical allocation rules \emph{all of which respect the same partial ordering within a single bidder's types, but may not respect any global ordering across all bidders' types}. 

Indeed, a stronger charcaterization is possible in the form of Theorem~\ref{thm:non-iidhierarchical}: for every feasible reduced form $\mathcal{R}$ there exists a \emph{global} total ordering of (bidder, type) pairs $\succ$, such that $\mathcal{R}$ can be induced by a distribution of hierarchical allocation rules that all respect $\succ$. We prove Theorem~\ref{thm:non-iidhierarchical} in Section~\ref{sec:stronger} using a combinatorial approach, and also show that the algebraic approach of Sections~\ref{sec:shaded} and~\ref{sec:asymmetrichierarchical} cannot possibly yield such a theorem.
\end{enumerate}}

Let us now proceed by first recalling Equation~\eqref{eq:improved feasibility condition non symmetric}, which states that a reduced form $\cR$ is feasible if and only if for all thresholds $x_1,\ldots, x_n$, the probability that $\cR$ awards the item to a bidder $i$ whose reported type $\tau_i$ satisfies $\pi_i(\tau_i) \geq x_i$ is at most the probability that such a type is reported. When~\eqref{eq:improved feasibility condition non symmetric} is violated at some choice of thresholds $\vec{x}$, we call these thresholds \emph{constricting}. Unfortunately, there are roughly $c^n$ relevant $\vec{x}$ to test, so testing each of them separately is computationally/analytically intractable. 

What we would really like is a more structured subset of Border constraints that are sufficient to check. To this end, we show that properly \emph{shading} interim allocation probabilities according to a bidder's type distribution allows interim probabilities to be compared across bidders at face value. 

\subsection{The Shaded Reduced Form}\label{sec:shaded}
In this section, we define our notion of a \emph{shaded reduced form}. We first provide the definition and main proposition, followed by some illustrative examples where analysis is greatly simplified by our approach.


\begin{definition}\label{def:virtual pi} For any $s_i(\cdot)$ such that $s_i(\tau) \in \left[\Pr[\pi_i(t_i) < \pi_i(\tau)], \Pr[\pi_i(t_i) \leq \pi_i(\tau)]\right]$ for all $i, \tau$, and any reduced form $\pi$, we define the corresponding {\em shaded reduced form $\hat{\pi}$} as follows: for all $i$ and types $\tau \in T$, $\hat{\pi}_i(\tau):=s_i(\tau) \cdot \pi_i(\tau)$.
\end{definition}

\begin{observation}\label{obs: one}For all bidders $i$ and any types $\tau,\tau'\in T$, $\hat{\pi}_{i}(\tau) \geq \hat{\pi}_{i}(\tau') \implies \pi_{i}(\tau) \geq \pi_{i}(\tau')$.
\end{observation}
\begin{proof} If $\hat{\pi}_{i}(\tau) \geq \hat{\pi}_{i}(\tau')$ but $\pi_{i}(\tau) < \pi_{i}(\tau')$, clearly $s_i(\tau)>s_i(\tau')$. On the other hand, $s_i(\tau)\leq \Pr\left[\pi_i(t_i)\leq \pi_i(\tau)\right]\leq \Pr\left[\pi_i(t_i)< \pi_i(\tau')\right]\leq s_i(\tau')$. Contradiction. \end{proof}

\begin{proposition}\label{prop: virtualsuffices}
Let $\pi$ be an infeasible reduced form, and $s_i(\cdot)$ be any shading satisfying $s_i(\tau) \in [\Pr[\pi_i(t_i) < \pi_i(\tau)], \Pr[\pi_i(t_i) \leq \pi_i(\tau)]]$ for all $i, \tau$. Then there exists a single threshold $x$ such that: $\sum_i \sum_{\tau_i | \hat{\pi}_i(\tau_i) \geq x} \pi_i(\tau_i) \cdot \Pr[t_i = \tau_i] > 1 - \prod_i \left(1-\Pr[\hat{\pi}_i(t_i) \geq x]\right)$.

In other words, for any valid shading of the reduced form, one can determine feasibility of a reduced form by checking Border's constraints where the threshold for the shaded reduced form is constant across all bidders.
\end{proposition}

\begin{proof} If a reduced form $\cR$ is infeasible, consider any maximal choice of constricting thresholds $x_1,\ldots, x_n$, i.e. a choice of $x_1,\ldots, x_n$ such that $(x_i+\delta,x_{-i})$ is not constricting for any $i, \delta > 0$. Now let ${(i,\tau)} \in \argmin_{j,\mu: \pi_j(\mu) \geq x_j} \{\hat{\pi}_{j}(\mu)\}$. Then by the maximality of $x_1,\ldots, x_n$, it must be the case that decreasing from $x_i +\delta$ to $x_i$ causes us to go from satisfying~\eqref{eq:improved feasibility condition non symmetric} to violating it, and therefore {$x_i=\pi_i(\tau)$} and this change must increase the LHS by more than it increases the RHS. The change in the LHS is easy to compute: Observation~\ref{obs: one} implies that we are simply including additional (bidder, type) pairs in our calculations, namely $(i, \tau)$ (and all $(i, \tau')$ with $\pi_i(\tau') = \pi_i(\tau)$). The change in the RHS is also easy to compute: we have increased the probability that some bidder $k$ exists with $\pi_k(t_k) \geq x_k$ by exactly the probability that all bidders $j \neq i$ have $\pi_j(t_j) < x_j$ and $\pi_i(t_i) = \pi_i(\tau)$. This therefore implies:

$$\Pr[\pi_i(t_i) = \pi_i(\tau)]\pi_{i}(\tau) > \Pr[\pi_i(t_i) = \pi_i(\tau)]\prod_{j \neq i}\Pr[\pi_j(t_j) < x_j] \iff \frac{\pi_{i}(\tau)}{\prod_{j \neq i}\Pr[\pi_j(t_j) < x_j]} > 1.$$

Now consider any other $\tau', k$, $\pi_k(\tau') < x_k$ and $\hat{\pi}_{k}(\tau') \geq \hat{\pi}_{i}(\tau)$. Observe first that we must have $k \neq i$, as Observation~\ref{obs: one} would otherwise imply $\pi_i(\tau') \geq x_i$. So we must have:
\begin{align*}
&\pi_{k}(\tau')\cdot \Pr[\pi_k(t_k) \leq \pi_{k}(\tau')] \geq \hat{\pi}_k(\tau') \geq \hat{\pi}_i(\tau) \geq \pi_{i}(\tau)\cdot \Pr[\pi_i(t_i) < \pi_{i}(\tau)]\quad {(\text{valid shading)}}\\
\Longrightarrow& \pi_{k}(\tau')\cdot \Pr[\pi_k(t_k) < x_k] \geq \pi_{i}(\tau)\cdot \Pr[\pi_i(t_i)< \pi_{i}(\tau)]\quad {(\text{because }\pi_k(\tau')<x_k)}\\
\Longleftrightarrow& \pi_{k}(\tau')\cdot\Pr[\pi_k(t_k) < x_k] \geq \pi_{i}(\tau)\cdot\Pr[\pi_i(t_i)< x_{i}]\quad {(\text{because } \pi_i(\tau) = x_i)}\\
\Longleftrightarrow& \frac{\pi_{k}(\tau')}{\prod_{j \neq k} \Pr[\pi_j(t_j) < x_j]} \geq \frac{\pi_{i}(\tau)}{\prod_{j \neq i}\Pr[\pi_j(t_j) < x_j]}.
\end{align*}
So by our choice of $(i,\tau)$ and the work above, we obtain:
$$\frac{\pi_{k}(\tau')}{\prod_{j \neq k}\Pr[\pi_j(t_j) < x_j]} > 1\iff \Pr[t_k = \tau']\pi_{k}(\tau') > \Pr[t_k = \tau']\prod_{j \neq k} \Pr[\pi_j(t_j) < x_j].$$

This inequality (combined with Observation~\ref{obs: one}) tells us that {we could lower $x_k$ to $\pi_k(\tau')$ and still have constricting thresholds.} In fact, it tells us something even stronger. If $y_1,\ldots,y_n$ are constricting thresholds with $y_j \leq x_j$ for all $j$, then we could decrease $y_k$ to $\pi_k(\tau')$ and still have constricting thresholds. This is because lowering $y_k$ to $\pi_k(\tau')$, causes the LHS of~\eqref{eq:improved feasibility condition non symmetric} to increase by $\sum_{\tau_k: \pi_k(\tau_k)\in[\pi_k(\tau'),y_k)}\Pr[t_k=\tau_k]\pi_k(\tau_k)\geq \Pr[\pi_k(t_k) \in\left[\pi_k(\tau'),y_k)\right]\pi_k(\tau')$. 
And the RHS increases by exactly {$\Pr\left[\pi_k(t_k) \in[\pi_k(\tau'),y_k)\right]$} times the probability that $\pi_j(t_j) < x_j$ (respectively, $y_j$) for all $j\neq k$. As we decrease from $x_j$ to $y_j$, this probability will clearly never increase.

So starting from any constricting thresholds $x_1,\ldots,x_n$ and a bidder type pair $(i, \tau)$ as above, {we can lower each bidder $k$'s threshold $x_k$ to the lowest $\pi_k(\tau')$ such that $\pi_k(\tau')\geq \hat{\pi}_{i}(\tau)$.} 
{The resulting thresholds remain constricting and have the desired form.}

\end{proof}

The proof of Theorem~\ref{thm: independent} now readily follows from Proposition~\ref{prop: virtualsuffices} and similar routine computation to Corollary~\ref{cor:algiid}. A complete proof of Theorem~\ref{thm: independent} appears in Appendix~\ref{app:proofs}. 
{We now proceed with a few examples to clearly illustrate the benefits of an improved characterization.}

\subsubsection{Some Illustrative Examples}\label{sec:examples}

\paragraph{Example One: Monomials.} 
Consider a reduced form where $T = [0,1]$ and each $\mathcal{D}_i = U(T)$ (the uniform distribution on $[0,1]$). Moreover, for all $i$ let $\pi_i(\tau) = \tau^{\alpha_i}$, for some constants (but not necessarily identical) $\alpha_i$. Using prior work, one could check all constraints of the form~\eqref{eq:improved feasibility condition non symmetric}, which is a multi-variate optimization problem: maximize $\sum_{i=1}^n \int_{y_i}^1 \tau^{\alpha_i} d\tau- (1-\prod_{i=1}^n y_i)$, over all $\vec{y} \in [0,1]^n$ (to map onto the variables used in Equation~\eqref{eq:improved feasibility condition non symmetric}, set $x_i = y_i^{\alpha_i}$). If the maximum happens to yield a value $\leq 0$, then all constraints are satisfied. If the maximum yields a value $>0$, then we have explicitly found a violated constraint. In turn, identifying the maximum requires considering second-order conditions at all local optima in addition to all points on the boundary, and is tedious.\footnote{Note that Che et al. extend Border's Theorem to continuous type spaces (the theorem statement is identical to that for discrete type spaces)~\cite{CheKM11}. Given this, it is easy to see that our proof of Proposition~\ref{prop: virtualsuffices} extends to continuous type spaces as well as-is (subject to some change in notation).}

With Theorem~\ref{thm: independent} in hand, we observe that we need not optimize over all settings of thresholds $\vec{y} \in [0,1]^n$. Rather, it suffices to only consider thresholds that are jointly parametrized via a single threshold $x \in [0,1]$ on the shaded reduced form. For all $x\in [0,1]$, the corresponding cutoffs $\vec{y}$ satisfy $y_i^{\alpha_i+1} = x$, for all $i$: This is because for any $\tau \in [0,1]$, $\Pr[\pi_i(t_i) \leq \pi_i(\tau)] = \tau$, and $\pi_i(\tau) = \tau^{\alpha_i}	$, so $\Pr[\pi_i(t_i) \leq \pi_i(\tau)] \cdot \pi_i(\tau) = \tau^{\alpha_i + 1}$. So we see that our problem now reduces to a single-variate optimization: maximize $\sum_{i=1}^n \int_{x^{1/(\alpha_i+1)}}^1 \tau^{\alpha_i}d\tau - (1-\prod_{i=1}^n x^{1/(\alpha_i+1)})$, over all $x \in [0,1]$. 


This integral happens to be extremely simple to evaluate, and our objective function is just $\sum_{i=1}^n \frac{1-x}{\alpha_i + 1} - \left(1-x^{(\sum_{i=1}^n 1/(\alpha_i+1))}\right)$. We can then take a derivative with respect to $x$, which is $\left(\sum_{i=1}^n \frac{1}{\alpha_i+1}\right)\cdot \left(x^{(\sum_{i=1}^n 1/(\alpha_i+1)-1)}-1\right)$. At this point, we can observe that if $\sum_{i=1}^n 1/(\alpha_i+1) > 1$, then the derivative is negative on the entire interval $[0,1)$, and therefore the maximum occurs at $x = 0$. At $x = 0$, the objective function evaluates to $\sum_i 1/(\alpha_i+1) - 1 > 0$, and therefore such reduced forms are infeasible. If instead, $\sum_i 1/(\alpha_i+1) \leq 1$, then the derivative is non-negative on the entire interval $[0,1)$, and therefore a maximum occurs at $x = 1$. At $x = 1$, the objective function evaluates to $0$, and therefore such reduced forms are feasible.

In conclusion, Theorem~\ref{thm: independent} combined with the single-variable optimization above provides a complete proof that reduced forms of the above form are feasible if and only if they promise at most one item in expectation ex ante, which occurs if and only if $\sum_{i=1}^n 1/(\alpha_i + 1) \leq 1$. 

\paragraph{Example Two: Uniform Distributions of Support Two.} Now, consider a reduced form where $T = \{H, L\}$, and each $\mathcal{D}_i = U(\{H,L\})$. Then if $\pi_i(H) \geq \pi_j(H)/2$ for all $i,j$, a valid shading sets $\hat{\pi}_i(H) = \min_j\{\pi_j(H)\}$, and $\hat{\pi}_i(L) = 0$ for all $i$. Proposition~\ref{prop: virtualsuffices} then guarantees that we only need to check two constraints: $\sum_{i=1}^n \pi_i(H)/2 \leq 1-1/2^n$, and $\sum_{i=1}^n \pi_i(H)/2 +\pi_i(L)/2 \leq 1$. On the other hand, using prior work would require checking $2^n$ equations of the form~\eqref{eq:improved feasibility condition non symmetric} (with some extra thought, this can be reduced to $n$ by appealing to the fact that the type distributions are iid. But getting all the way down to $2$ requires reasoning \`{a} la Proposition~\ref{prop: virtualsuffices}). 

Similarly, if each $\mathcal{D}_i$ assigns probability $p_i$ to $L$, and $\pi_i(H) \geq p_j\pi_j(H)$ for all $i,j$, a valid shading sets $\hat{\pi}_i(H) = \min_j\{\pi_j(H)\}$, and $\hat{\pi}_i(L) = 0$ for all $i$. Proposition~\ref{prop: virtualsuffices} again guarantees that we only need to check two constraints: $\sum_{i=1}^n (1-p_i) \cdot \pi_i(H) \leq 1- \prod_{i=1}^n p_i$, and $\sum_{i=1}^n (1-p_i) \cdot \pi_i(H) + p_i \cdot \pi_i(L) \leq 1$. On the other hand, using prior work would again require checking $2^n$ equations of the form~\eqref{eq:improved feasibility condition non symmetric} (and this time it is not obvious how to check any fewer).

\paragraph{Example Three: A Correct Ordering is Necessary.} The above two examples clearly illustrate why a tighter characterization is helpful, but it is not a priori clear that the even simpler approach of just considering thresholds where $x_i = x_j$ for all $i,j$ fails. To see that indeed this approach fails, consider the following example with two bidders, and two types per bidder, $H$ and $L$. For bidder $1$, we set $\Pr[t_1 = H] = 1/8$, $\Pr[t_1 = L] = 7/8$, $\pi_{1}(H) = 5/8$, $\pi_{1}(L) = 0$. For bidder $2$, we set $\Pr[t_2 = H] = 1/2$, $\Pr[t_2 = L] = 1/2$, $\pi_{2}(H) = 1$, $\pi_{2}(L) = 3/4$. 

This reduced form is infeasible. Indeed, observe that bidder $2$ must always receive the item whenever $\tau_2 = H$, which happens with probability $1/2$. So if we have $\pi_{2}(H) = 1$, we cannot also have $\pi_{1}(H) > 1/2$. So~\eqref{eq:improved feasibility condition non symmetric} is violated when $x_1 = 5/8$ and $x_2 = 1$. 

However, one can check that Border's conditions are satisfied whenever $x_1 = x_2$.\footnote{There are four inequalities of this form: $\Pr[t_2 = H] \cdot \pi_2(H) = 1/2 \leq 1 - \pi_2(L)$, $\Pr[t_2 = H] \cdot \pi_2(H) + \Pr[t_2 = L] \cdot \pi_2(L) = 7/8 \leq 1$, $\Pr[t_2 = H] \cdot \pi_2(H) + \Pr[t_2 = L] \cdot \pi_2(L) + \Pr[t_1 = H] \cdot \pi_1(H) = 61/64 \leq 1$, and $\Pr[t_2 = H] \cdot \pi_2(H) + \Pr[t_2 = L] \cdot \pi_2(L) + \Pr[t_1 = H] \cdot \pi_1(H)  + \Pr[t_1 = L] \cdot \pi_1(L) = 61/64 \leq 1$, all of which are satisfied.} Essentially the problem is that to find constricting thresholds, we need $x_1$ to be small enough to include $(1,H)$, yet $x_2$ big enough to exclude $(2,L)$, which is not possible when $x_1 = x_2$.



\paragraph{Examples: Summary.} Examples One and Two demonstrate the usefulness of a tighter characterization. On the other hand, Example Three shows that such a tighter characterization must be constructed carefully. Indeed, the shaded reduced form is exactly what is necessary to determine whether a reduced form is feasible or not. 

\subsection{Asymmetric Hierarchical Allocation Rules}\label{sec:asymmetrichierarchical}
Theorem~\ref{thm: independent} tightens the necessary and sufficient conditions of Border's theorem in a way that allows for computationally efficient determination of the feasibility of reduced forms. We proceed by examining which ex-post allocation rules are neccessary to induce all feasible reduced forms, similar to Theorem~\ref{thm:iidhierarchical}, first formally stating the definition of a hierarchical allocation rule in asymmetric settings.

\begin{definition}\label{def:hierarchicalas}
A \textbf{hierarchical allocation rule} consists of a weak total ordering $\succeq$ on $T \times [n] \cup \{(0,\bot)\}$. On reported types $(\tau_1,\ldots,\tau_n)$, the allocation rule computes the subset of indices $\{i|(i, \tau_i) \succeq (j, \tau_j), \forall j\}$, then selects a uniformly random index $i \in \cW$. If $i > 0$, the item is allocated to bidder $i$. If $i = 0$, the item is not allocated.

We say that a hierarchical allocation rule $\succeq$ for non-identical bidders is \emph{partially-ordered} {with respect to $\cR$} if for all $i$ and $\tau,\tau' \in T$, $\pi_i(\tau) \geq \pi_i(\tau') \Rightarrow (i,\tau) \succeq (i,\tau')$. We say that a hierarchical allocation rule is \emph{strict} if for all bidders $i\neq j$ and types $\tau, \tau' \in T$: $(i,\tau)\succeq (j,\tau') \wedge (i,\tau) {\succeq} (0,\bot) \Rightarrow (j,\tau') \not \succeq (i,\tau)$, and $(i,\tau) \succeq (0,\bot) \Rightarrow (0,\bot) \not \succeq (i,\tau)$ (i.e. $|\cW| = 1$ on all inputs).
\end{definition}

Similar to the symmetric case, a simple counting argument shows that every feasible reduced form can be implemented as a distribution over strict, partially-ordered hierarchical allocation rules. The proof for the asymmetric case follows the same outline, but requires one additional technical lemma whose proof is deferred to the Appendix~\ref{app:proofs}.

{\begin{theorem}\label{thm: independentcorners} {(implied by~\cite{ManelliV10})} Every feasible reduced form $\cR$ lies in a $cn$-dimensional polytope whose corners are all strict, partially-ordered w.r.t. $\cR$ hierarchical allocation rules. Furthermore, there is a distribution over at most $cn+1$ hierarchical allocation rules, all strict and partially-ordered w.r.t. $\cR$, that induces $\cR$. 
\end{theorem}}

\begin{proof}
For ease of notation, relabel all types in $T$ (differently for each bidder) so that $\pi_i(\tau_{i,1})\geq \ldots \geq\pi_i(\tau_{i,c})$ for all $i${, and so that $\Pr[t_i = \tau_{i,j}] = 0 \Rightarrow \Pr[t_i = \tau_{i, k}] = 0$ for all $k > j$ (i.e. since $\pi_i(\tau) = 0$ for all $\tau$ such that $\Pr[t_i = \tau] = 0$, we are free to put them at the end of the list). Let $c_i \leq c$ denote the number of types $\tau \in T$ such that $\Pr[t_i = \tau] > 0$}. Again for notational convenience, denote by $\pi_i(\tau_{i,c+1}) = 0$. Let $S_i$ denote the set of $j$ such that $\pi_i(\tau_{i,j}) = \pi_i(\tau_{i,j+1})$. {Consider the closed, convex polytope $P\subseteq [0,1]^{cn}$ specified by the following constraints.} 
\begin{align}
\tilde{\pi}_i(\tau_{i,j}) &= \tilde{\pi}_i(\tau_{i,j+1}) &\forall i \in [n], j \in S_i \label{eq:matt}\\
\tilde{\pi}_i(\tau_{i,j}) &\geq \tilde{\pi}_i(\tau_{i,j+1}) &\forall i \in [n],j \in [c]-S_i \label{eq:costas}\\
\sum_i \sum_{j < z_i}\Pr[t_{i}=\tau_{i,j}] \tilde{\pi}_i(\tau_{i,j}) &\leq 1 - \prod_i\left(1-\sum_{j < z_i}\Pr[t_i = \tau_{i,j}]\right) &\forall z_1,\ldots,z_n \in [c+1]\label{eq: wasteful}
\end{align}
where for notational convenience we denote $\tilde{\pi}_i(\tau_{i,c+1}) = 0$ (again, $\tilde{\pi}_i(\tau_{i,c+1})$ is not a free variable). In fact, we can also replace~\eqref{eq: wasteful} with:{
\begin{align}
\sum_i \sum_{j <z_i}\Pr[t_{i}=\tau_{i,j}] \tilde{\pi}_i(\tau_{i,j}) \leq &1- \prod_i \left(1-\sum_{j <z_i} \Pr[t_i = \tau_{i,j}]\right) &\forall z_1\in [c_1],\ldots,z_n \in [c_n ] \label{eq: simpler}\\
\sum_i \sum_{j < c_i+1}\Pr[t_{i}=\tau_{i,j}] \tilde{\pi}_i(\tau_{i,j}) &\leq 1 &\label{eq: better}\\
\tilde{\pi}_i(\tau_{i,j}) &= 0  &\forall i \in [n], j > c_i \label{eq:easy}
\end{align}}
In the above replacement, we are observing that if \eqref{eq: better} holds, then so does \eqref{eq: wasteful} for any case where at least one $i$ has $z_i = c_i+1$, {as the left-hand side of all such inequalities is upper bounded by the left-hand side of Equation~\eqref{eq: better}, and the right-hand side of all such inequalities is also $1$}. In addition, \eqref{eq: simpler} covers all other cases{, and Equation~\eqref{eq:easy} just states that $\tilde{\pi}_i(\tau) = 0$ whenever $\Pr[t_i = \tau] = 0$}. {We know from \cite{Border07,CheKM11} that $\cR$ is feasible if and only if $\pi\in P$.}

We proceed to show that any corner of this polytope is the reduced form of a strict hierarchical allocation rule. Note that any corner corresponds to a set of at least $cn$ tight inequalities from~\eqref{eq:matt}, \eqref{eq:costas}, \eqref{eq: simpler},~\eqref{eq: better}, or~\eqref{eq:easy} whose tightness determine a unique solution. Similarly to Theorem~\ref{thm:iidhierarchical}, we examine the structure of what constraints can be simultaneously tight. We again refer to the inequalities in~\eqref{eq:matt}/\eqref{eq:costas}/\eqref{eq:easy} as the $(i,j)^{th}$ monotonicity constraints, and those in~\eqref{eq: simpler}/\eqref{eq: better} as the $\vec{z}$    Border constraints (where $\vec{z}$ indexes the constraint as above). 

We first argue that the tight Border constraints must be \emph{nested}. That is, if the Border constraint is tight for both $\vec{z}$ and $\vec{w}$, then we must either have $z_i \leq w_i$ for all $i$, or $w_i \leq z_i$ for all $i$. Assume for contradiction that this is not the case. Then there exists some bidders $j, k$ such that {$w_j > z_j$ and $z_k > w_k$}. Clearly, we have $z_i < c_i+1$ (respectively, $w_i < c_i+1$) for all $i$, because otherwise we would necessarily have $z_i = c_i+1$ for all $i$ (respectively, $w_i = c_i+1$ for all $i$) - the only such constraint with any $z_i = c_i+1$ is~\eqref{eq: better}.  Consider now the type profile where bidder $j$ has type $\tau_{j,z_j}$, bidder $k$ has type $\tau_{k,w_k}$, and every other bidder $i$ has type $\tau_{i, c_i}$ {(note that this profile indeed arises with non-zero probability by definition of $c_i$)}. Now, any ex-post allocation rule that induces a feasible reduced form whose Border constraint at {$\vec{w}$} 
 is tight \emph{must} award the item to bidder $j$ on this profile, as every other bidder $i$'s type has index at least 
 {$w_i$}. Similarly, any ex-post allocation rule that induces a feasible reduced form whose Border constraint at 
 {$\vec{z}$} is tight \emph{must} award the item to bidder $k$ on this same profile, as every other bidder $i$'s type has index at least 
 {$z_i$}. Clearly, no feasible ex-post allocation rule can award the item to both bidders, so no feasible reduced form can have both Border constraints be tight. 

Now that we know that for any corner $\tilde{\pi}$ of $P$ the tight Border constraints are nested, we (suggestively) define the relation:
$$(i,\tau_{i,j}) \succeq (k, \tau_{k,\ell}) \Leftrightarrow \left|\{\vec{z}\ | z_i > j \wedge \text{ Border($\vec{z}$) is tight}\}\right| \geq \left|\left\{\vec{z}\ | z_k > \ell \wedge \text{ Border($\vec{z}$) is tight}\right\}\right|,$$
$$(i,\tau_{i,j}) \succeq (0,\bot) \Leftrightarrow \tilde{\pi}_i(\tau_{i,j}) > 0,~~~ (0,\bot) \succeq (i,\tau_{i,j}) \Leftrightarrow \tilde{\pi}_i(\tau_{i,j}) = 0.$$

Next, we reason about what monotonicity constraints can possibly be tight simultaneously with a nested set of Border constraints that define $\succeq$ with the following lemma (whose proof appears in Appendix~\ref{app:proofs}). 

\begin{lemma}\label{lem:tightmono}
For $\succeq$ defined as above for some $\tilde{\pi} \in P$, the $(i,j)^{th}$ monotonicity constraint can be tight and linearly independent of the tight Border constraints only if $(i,\tau_{i,j+1}) \succeq (i,\tau_{i,j})$. 
\end{lemma}

The remainder of the proof is just a counting argument. Define an equivalence relation $(i,\tau) \sim (j,\tau') \Leftrightarrow (i,\tau) \succeq (j,\tau') \succeq (i,\tau)$. Then because the tight Border constraints are nested, the number of tight Border constraints is exactly the number of equivalence classes under $\sim$ among (type, bidder) pairs $\succeq (0,\bot)$. {It is also now clear, from Lemma~\ref{lem:tightmono}, that a monotonicity constraint can be tight and linearly independent of the tight Border constraints only if it is between two types $(i,\tau_{i,j})$ and $(i,\tau_{i,j+1})$ of the same bidder $i$ and $(i,\tau_{i,j}) \sim (i,\tau_{i,j+1})$. Therefore, there are no tight monotonicity constraints across equivalence classes, and the number of tight monotonicity constraints in each equivalence class is at most the number of types in that class minus one. Moreover, the number of tight monotonicity constraints in each equivalence class can only be equal to the number of types in that class minus one if all types in that class are from the same bidder.} We simply observe that if $(i,\tau) \sim (j,\tau') \succeq (0,\bot)$ for any $i \neq j$, that the above counting shows we can't possibly have $cn$ tight linearly independent constraints. So we may conclude that $(i,\tau) \not\sim (j,\tau')$ for any $(i,\tau),(j,\tau') \succeq (0,\bot)$. 

Finally, we now want to conclude that the hierarchical allocation rule defined by $\succeq$ induces the proposed corner $\tilde{\pi}$. We make one slight modification to $\succeq$ to fit exactly Definition~\ref{def:hierarchicalas}, and merge adjacent equivalence classes that contain types from the same bidder. More formally, if there exist two types $(i,\tau_{i,j})$, $(i,\tau_{i,j+1})\succeq (0,\bot)$ that do not lie in the same equivalence class and no bidder $k\neq i$ has a type $\tau$ with $(i,\tau_{i,j})\succeq (k,\tau)\succeq (i,\tau_{i,j+1})$, we merge the two equivalence classes by modifying $\succeq $ so that $(i,\tau_{i,j+1})\succeq(i,\tau_{i,j})$ (but keeping $\succeq$ otherwise the same).\footnote{This adjustment is technically necessary to claim that when $\tilde{\pi}_i(\tau_{i,j}) = \tilde{\pi}_i(\tau_{i,j+1})$, we have $\tau_{i,j+1} \succeq \tau_{i,j}$. This adjustment doesn't affect the implementation of the hierarchical allocation rule according to $\succeq$ at all, since there is only ever one type per bidder present at the auction.} Now, it is clear that $\succeq$ is a weak total-ordering that is strict (because $(i,\tau) \not \sim (j,\tau')$ for any $(i,\tau),(j,\tau') \succeq (0,\bot)$) and partially-ordered w.r.t. to $\tilde{\pi}$, and the hierarchical allocation rule corresponding to $\succeq$ uniquely implements the corner $\tilde{\pi}$ (because the tight Border constraints uniquely determine a winner on every possible type profile, exactly the strongest type according to $\succeq$). 



The final sentence of the theorem statement is again a consequence of Carath\'{e}odory's Theorem and Observation~\ref{obs:distribution}.
\end{proof}

Now that we know that every corner of the polytope can be implemented as a strict, partially-ordered w.r.t. $\mathcal{R}$ hierarchical allocation rule, we want to ensure that the hierarchy $\succeq$ can be found computationally efficiently.

\begin{lemma}\label{lem:corner to allocation}
	Let $\tilde{\pi}$ be any corner of $P$ and $\succeq$ be the strict and partially-ordered w.r.t. $\mathcal{R}$ hierarchical allocation rule that implements $\tilde{\pi}$. Then the shaded reduced form defined as $\hat{\tilde{\pi}}_i(\tau)=\tilde{\pi}_i(\tau)\cdot \Pr\left[\tilde{\pi}_i(t_i)\leq \tilde{\pi}_i(\tau)\right]$ respects $\succeq$. Specifically, for any two types $(i,\tau)$ and $(j,\tau')$, $(i,\tau)\succeq (j,\tau')\iff \hat{\pi}_i(\tau)\geq \hat{\pi}_j(\tau')$. Therefore, given $\tilde{\pi}$ we can construct the ordering $\succeq$ in time $O(cn\log (cn))$.
\end{lemma}	
\begin{proof}
First, observe that for any type $(i,\tau)$, $\tilde{\pi}_i(\tau)= \prod_{k\neq i} \Pr[(i,\tau)\succeq (k,t_k)]$ and $\Pr\left[\tilde{\pi}_i(t_i)\leq \tilde{\pi}_i(\tau)\right]=\Pr[(i,\tau)\succeq (i,t_i)]$.
Therefore, $\hat{\tilde{\pi}}_i(\tau)= \prod_{k=1}^n \Pr[(i,\tau)\succeq (k,t_k)]$. We may therefore immediately conclude that $(i,\tau) \succeq (j,\tau') \Leftrightarrow \hat{\tilde{\pi}}_i(\tau) \geq \hat{\tilde{\pi}}_j(\tau')$. So in order to find the ordering $\succeq$, we only need to compute the shaded reduced form and sort its components, which can clearly be done in time $O(cn\log(cn))$.
\end{proof}

And now, we can draw the main conclusion of this section: given as input any reduced form, we can computationally efficiently determine whether or not it is feasible. If it is feasible, we can computationally efficiently output an implementation.

\begin{corollary}\label{cor:grand theorem independent} 
Given an asymmetric reduced form $\cR$, one can determine if it is feasible or find a hyperplane separating it from the set of feasible reduced forms, in time {$O(cn\log(cn))$}. If $\cR$ is feasible, a succinct description of an allocation rule implementing $\cR$ can be found in time polynomial in $c$ and $n$. {The output allocation rule is a distribution over at most $cn +1$ hierarchical allocation rules, all strict and partially-ordered w.r.t. $\cR$}.
\end{corollary}

\begin{proof}
We first observe that Theorem~\ref{thm: independent} provides an algorithm that determines if $\cR$ is feasible, or provides a hyperplane separating it from the space of feasible reduced forms (the violated Border constraint) that runs in time $O(cn\log(cn))$. We now have to describe how to efficiently find an ex-post allocation rule implementing a reduced form $\cR$ that is feasible. Theorem~\ref{thm: independentcorners} implies that $\pi$ lies inside a $cn$-dimensional polytope, $P$, whose corners are the reduced forms of the strict, partially-ordered w.r.t. $\cR$ hierarchical allocation rules. Theorem~\ref{thm: independent} provides a separation oracle for $P$, so Theorem~\ref{thm:GLS} implies that we can decompose $\pi$ into a convex combination of corners of $P$ in time polynomial in $cn$ and $cn\log(cn)$ (resulting in a runtime polynomial in both $c$ and $n$). Lemma~\ref{lem:corner to allocation} shows how to implement a hierarchical allocation rule in time $O(cn\log (cn))$ given a corner.  Observation~\ref{obs:distribution} completes the proof.
\end{proof}

\subsection{A Tighter Characterization Result}\label{sec:stronger}
{
Let's first briefly recall the goal of this section. Theorem~\ref{thm: independentcorners} provides a nice characterization result: every feasible reduced form can be induced by a distribution over hierarchical allocation rules, all of which respect the same partial ordering within a single bidder's types, but may not respect any global ordering across all bidders' types. The purpose of this section is to prove Theorem~\ref{thm:non-iidhierarchical} and show that in fact the distribution over hierarchical allocation rules may be taken to respect the same global ordering over all bidders' types. 

{At this point, we note that it would be great if the shaded reduced form provided an easy way to extend Theorem~\ref{thm:iidhierarchical} to the asymmetric setting. Specifically, we can say that a hierarchical allocation rule $\succeq$ is \emph{shaded-ordered} if {$\hat{\pi}_i(\tau) \geq \hat{\pi}_j(\tau') \Rightarrow (i,\tau) \succeq (j,\tau')$, and hope it is the case that every feasible reduced form can be induced by a distribution over shaded-ordered hierarchical allocation rules. Unfortunately, although the shaded reduced form provides a nice structural theorem about feasible reduced forms and a near-linear time algorithm for determining feasibility, the following example shows that distributions over shaded-ordered hierarchical allocation rules are not sufficient to implement every feasible reduced form when the bidders are non-i.i.d.} For completeness, we rule out all possible shadings.

\begin{proposition}\label{obs: novirtual} There exist feasible reduced forms that cannot be induced by distributions over shaded-ordered hierarchical allocation rules.
\end{proposition}
\begin{proof} 
Consider the following example with two bidders and two types, with $\epsilon <1/4$. Bidder one has $\Pr[t_1 =H] = 1-\epsilon^2, \Pr[t_1 = L] = \epsilon^2$, $\pi_1(H) = 1-\epsilon^2$ and $\pi_1(L) = 1-2\epsilon^2$. Bidder two has $\Pr[t_2 = H] = \epsilon, \Pr[t_2 = L] = 1-\epsilon$, $\pi_2(H) = \epsilon$ and $\pi_2(L) = 0$. 

Then for any shading, we have $\hat{\pi}_1(H) \geq (1-\epsilon^2)^2 >\epsilon \geq \hat{\pi}_2(H) \geq \epsilon\cdot (1-\epsilon) > \epsilon^2 \cdot (1-2\epsilon^2) \geq \hat{\pi}_1(L)$. So any shaded-ordered reduced form necessarily has $(2,H) \succeq (1,L)$, and therefore cannot possibly have $\pi_1(L) > 1-\epsilon/2 \geq 1-2\epsilon^2$. 

Observe also that this reduced form is clearly feasible: consider the allocation rule that awards the item to bidder one whenever $t_2 = L$, awards the item to bidder one with probability $1-\epsilon$ when $t_1 = t_2 = H$ (and bidder two otherwise), and awards the item to bidder one with probability $1-2\epsilon$, to bidder two with probability $\epsilon$ when $t_1 = L, t_2 = H$ (and throws the item away otherwise). Then bidder one receives the item with probability $1-\epsilon^2$ when $t_1 = H$, with probability $1-2\epsilon^2$ when $t_1 = L$, and bidder two receives the item with probability $\epsilon$ when her type is $H$, and $0$ otherwise. 
\end{proof}




{In light of Proposition~\ref{obs: novirtual}, it seems that geometric techniques will not get us all the way to a proof of Theorem~\ref{thm:non-iidhierarchical} (which provides a global ordering instead of just a partial ordering as in Theorem~\ref{thm: independentcorners}), so our proof below appeals more to analytical tools. Throughout the proof, we will use the term $\succ$-ordered hierarchical allocation rule to denote a hierarchical allocation rule corresponding to some weak total ordering $\succeq$ that irons the strict total ordering $\succ$.
\smallskip

\begin{prevproof}{Theorem}{thm:non-iidhierarchical}
{The high-level approach is to find a strict total ordering, $\succ$, and an allocation rule that is a distribution over $\succ$-ordered hierarchical allocation rules, $M$, that is ``closest'' to inducing $\cR$ by some measure (over all $\succ, M$). We will then argue that if $M$ does not already induce $\cR$, then the fact that we cannot improve $M$ witnesses a violated Border constraint.}

We first formally introduce a dummy bidder $0$ with $\Pr[t_0 = \bot] = 1$ and $\pi_0(\bot) = 1 - \sum_{i > 0}\sum_{\tau_i} \Pr[t_i = \tau_i] \pi_i(\tau_i)$. With this addition, we now have $\sum_{i \geq 0} \sum_{\tau_i} \Pr[t_i = \tau_i] \pi_i(\tau_i) = 1$. So if we find a feasible allocation rule $M$ whose reduced form ${\pi}^M$ satisfies ${\pi}^M_i(\tau) \geq \pi_i(\tau)$ for all $i \geq 0, \tau$, then we must have ${\pi}^M = \pi$, and $M$ induces $\cR$. 

Let $\succ$ be a strict total ordering over all possible types that respects all the per-bidder partial orderings induced by $\pi$. Namely,  for all $i$, if $(i,\tau)\succ(i,\tau')$, then $\pi_i(\tau)\geq\pi_i(\tau')$. Define the unhappiness $F_{\succ}(M)$ of a distribution over $\succ$-ordered hierarchical allocation rules, $M$ (with reduced form $\pi^M$), as follows:
$$F_{\succ}(M)=\max_{i\geq 0,\tau\in T}(\pi_i(\tau)-\pi^M_i(\tau)).$$
{$F_{\succ}$ can be viewed as a continuous function over a compact set: There are finitely many $\succ$-ordered hierarchical allocation rules, so their convex hull is a compact set (and exactly the space of distributions over $\succ$-ordered hierarchical allocation rules). Each function $\pi_i(\tau) - \pi^M_i(\tau)$ is linear in this space (and therefore continuous), and the maximum of continuous functions is continuous. Hence, $F_{\succ}$ achieves its infimum.} Let then $M^{\succ} \in \argmin_{M} F_{\succ}(M)$ {(where the minimization is over all distributions over $\succ$-ordered hierarchical allocation rules)} and define the set $S_{\succ}$ to be the set of maximally unhappy types under $M^{\succ}$; formally, $S_{\succ}=\argmax_{i,\tau}\{\pi_i(\tau)-\pi^{M^{\succ}}_i(\tau)\}$. {If for some $\succ$ there are several minimizers $M^{\succ}$, choose one that minimizes $|S_{\succ}|$.} Now, let $MO$ (stands for Minimal Orderings) be the set of the orderings $\succ$ that minimize $F_{\succ}(M^\succ)$, further refined to only contain $\succ \in MO$ that also minimizing $|S_\succ|$. Formally, first set $MO=\argmin_{\succ}\{ F_{\succ}(M^\succ)\}$ and then {refine $MO$ as $MO_{\rm new} = \argmin_{\succ \in MO} \{|S_\succ|\}$. We drop the subscript ``${\rm new}$'' for the rest of the proof.}

From now on, we call a (bidder, type) pair $(i,\tau)$ {\em happy} if $\pi^M_i(\tau)\geq \pi_i(\tau)$, otherwise we call $(i,\tau)$ {\em unhappy}. Intuitively, here is what we have already done: For every ordering $\succ$, we have found a distribution over $\succ$-ordered hierarchical allocation rules $M^\succ$ that minimizes the maximal unhappiness {and subject to this, the number of maximally unhappy types}. We then choose from these $(\succ, M^{\succ})$ pairs those that minimize 
{the maximal unhappiness, and subject to this,} the number of maximally unhappy types. We have made these definitions because we want to eventually show that there is an ordering $\succ$, such that $F_{\succ}(M^\succ) = 0$, and it is natural to start with the ordering that is ``closest'' to satisfying this property. 

What we will show next is that, if $\exists \succ \in MO$ that does not make every (bidder, type) pair happy, then there also exists some $\succ' \in MO$, such that $F_{\succ'}(M^{\succ'})=F_{\succ}(M^\succ)$, $|S_{\succ'}|=|S_{\succ}|$, and $(i,\tau) \succ' (j,\tau')$ for all $(i,\tau) \in S_{\succ'}, (j,\tau') \notin S_{\succ'}$. In other words, only the top $|S_{\succ'}|$ types in $\succ'$ are maximally unhappy. From here, we will show that because $\succ' \in MO$, that $S_{\succ'}$ is a constricting set for $\cR$, contradicting its feasibility. We begin by showing the existence of $\succ'$, beginning with an arbitrary $\succ \in MO$. 

Before we begin, we introduce some terminology. We say that two (bidder, type) pairs $(i,\tau), (j,\tau')$ are \emph{adjacent} if $(i,\tau) \succ (k,\tau'') \Leftrightarrow (j,\tau') \succ (k,\tau'')$ for all $(k,\tau'') \notin \{(i,\tau),(j,\tau')\}$. For any $\succeq$, we also define an equivalence relation $\sim_\succeq$ with $(i,\tau) \sim_\succeq {(j,\tau')} \Leftrightarrow (i,\tau) \succeq (j,\tau') \wedge (j,\tau') \succeq (i,\tau)$. Finally, we say that there is a \emph{cut} between two adjacent types $(i,\tau)$ and $(j,\tau')$ in $\succeq$ if $(i,\tau) \succ (j,\tau')$ and $(j,\tau') \not \succeq (i,\tau)$. When we talk about adding a cut below $(i,\tau)$, we mean modifying $\succeq$ so that $(j,\tau') \not \succeq (i,\tau)$ for all $(i,\tau) \succ (j,\tau')$ (but otherwise keeping $\succeq$ the same). When we talk about removing a cut between two equivalence classes $A$ and $B$, we mean modifying $\succeq$ so that $(i,\tau) \sim_{\succeq} (j,\tau')$ for all $(i,\tau), (j,\tau') \in A \cup B$ (but otherwise keeping $\succeq$ the same).

Now, if $S_\succ$ is not the highest $|S_\succ|$ (bidder, type) pairs, let $(i,\tau)$ be the maximal element under $\succ$ in $S_\succ$ such that there exists some $(k,\tau'') \notin S_\succ$ with $(k,\tau'') \succ (i,\tau)$. Then the adjacent (bidder, type) pair $(j,\tau')$ with $(j,\tau') \succ (i,\tau)$ is necessarily $\notin S_\succ$. We proceed to show that we can change $\succ$ to swap $(i,\tau) \succ (j,\tau')$ (keeping $M$, $S_\succ$ and $F_\succ(M)$ as-is). We can repeat these swaps iteratively and they will terminate in the $\succ'$ we want (with $S_{\succ'}$ equal to the first $|S_{\succ'}|$ (bidder, type) pairs). 

We now proceed with a case analysis, for fixed $(j,\tau') \notin S_\succ$, $(i,\tau) \in S_\succ$, $(j,\tau') \succ (i,\tau)$ and $(j,\tau'),(i,\tau)$ adjacent.

\begin{itemize}
\item \textbf{Case 1:} $i=j$.

Since $\succ$ is a linear extension of the bidder's own ordering, we must have $\pi_i(\tau')\geq \pi_i(\tau)$, but we know that 
$$\pi_i(\tau')-\pi^{M^\succ}_i(\tau')<\pi_i(\tau)-\pi^{M^\succ}_i(\tau),$$
 thus $\pi^{M^\succ}_i(\tau')>\pi^{M^\succ}_i(\tau)\geq 0$. In any hierarchical mechanism $\succeq$, if there is no cut between $(i,\tau')$ and $(i,\tau)$, then they would receive the item with the same probability. Therefore, there must exist some $\succeq$ in the support of $M^\succ$ with a cut below $(i,\tau')$, and in which $(i,\tau')$ gets the item with non-zero probability. We modify $M^\succ$ {by modifying the hierarchical allocation rules $\succeq$ in its support as follows.} 
 
 Let $\succeq$ be a hierarchical allocation rule in the support of $M^\succ$. If there is no cut below $(i,\tau')$, we do nothing. If all (bidder, type) pairs equivalent to $(i,\tau')$ \emph{and} those equivalent to $(i,\tau)$ are of bidder $i$, we remove the cut below $(i,\tau')$. This does not affect the allocation probabilities at all, because it was impossible for two types equivalent to either $(i,\tau')$ or $(i,\tau)$ to show up together anyway. {So after this ``modification,'' we haven't changed $M^\succ$ at all, meaning that there must still exist some $\succeq$ in the support of $M^\succ$ with a cut below $(i,\tau')$, and in which $(i,\tau')$ gets the item with non-zero probability, and clearly it is not one of the allocation rules we just modified by removing the cut below $(i,\tau')$.} For such an $\succeq$, there is at least one (bidder, type) pair with bidder $\neq i$ equivalent to $(i,\tau')$ or $(i,\tau)$. {We distinguish two sub-cases:} 
\begin{itemize}

\item Every bidder $k \neq i$ has at least one type $\tau_k$ such that $(i,\tau) \succeq (k,\tau_k)$ (in other words, every (bidder, type) pair equivalent to $(i,\tau)$ wins the item with non-zero probability). Consider now moving the cut from below $(i,\tau')$ to right above $(i,\tau')$. Clearly, $(i,\tau')$ will be less happy if we do this. Every (bidder, type) pair with bidder $\neq i$ that was formerly equivalent to $(i,\tau')$ will be strictly happier, as now they do not have to share the item with $(i,\tau')$, whereas previously they did with positive probability. Every (bidder, type) pair with bidder $\neq i$ that is equivalent to $(i,\tau)$ will be strictly happier, as they now get to share the item with $(i,\tau')$, whereas previously they always lost to $(i,\tau')$. It is also clear to see that all $(i,\tau'')$, for $\tau'' \neq \tau'$ are unaffected by this change. So in particular $(i,\tau)$ is unaffected. 

Consider instead moving the cut from below $(i,\tau')$ to right below $(i,\tau)$. Then {$(i,\tau)$ is clearly strictly happier}, every (bidder, type) pair with bidder $\neq i$ that was formerly equivalent to $(i,\tau)$ is less happy than before (as they now don't get to share with $(i,\tau)$), every (bidder, type) pair with bidder $\neq i$ that is equivalent to $(i,\tau')$ is also less happy than before (because now they have to share with $(i,\tau')$), and all $(i,\tau'')$, for $\tau'' \neq \tau$ are not affected by the change. 

To summarize, we have argued that when we move the cut from below $(i,\tau')$ to just below $(i,\tau)$, $(i,\tau)$ becomes strictly happier, and every (bidder, type) pair that becomes less happy by this change becomes strictly happier if we instead move the cut to just above $(i,\tau')$ instead. Also, $(i,\tau)$ is unaffected by moving the cut to just above $(i,\tau)$. So with a tiny probability $\epsilon$, move the cut from below $(i,\tau')$ to just above $(i,\tau')$, whenever $\succeq$ is sampled from $M^{\succ}$. This makes all of the (bidder, type) pairs with bidder $\neq i$ that were equivalent to either $(i,\tau')$ or $(i,\tau)$ strictly happier. With a tinier probability $\delta$, move the cut from below $(i,\tau')$ to below $(i,\tau)$, whenever $\succeq$ is sampled from $M^{\succ}$. Choose $\epsilon$ to be small enough that we don't make $(i,\tau')$ maximally unhappy, and choose $\delta$ to be small enough so that we don't make any (type, bidder) pairs besides $(i,\tau')$ less happy than they were {in $\succeq$}. Then we have strictly increased the happiness of $(i,\tau)$ without making $(i,\tau')$ maximally unhappy, or decreasing the happiness of any other (bidder, type) pairs. Therefore, we have reduced $|S_\succ|$, contradicting the choice of $M^\succ$. 

\item If there is a bidder $k$ such that $(i,\tau) \not \succeq (k,\tau_k)$ for all $\tau_k$, (call such bidders {\em high}), then no (bidder, type) pair equivalent to $(i,\tau)$ can possibly win the item. We also know {that every high bidder} $k$ has at least one type $\tau_k$ such that $(k,\tau_k) \sim_\succeq (i,\tau')$ by our choice of $\succeq$ {(otherwise $(i,\tau')$ would get the item with probability $0$)}. Now we can basically use the same argument as above. The only difference is that when we move the cut to just above $(i,\tau')$ or just below $(i,\tau)$, (bidder, type) pairs formerly equivalent to $(i,\tau)$ (other than $(i,\tau)$ itself) will remain unaffected. But {since every high bidder $k$} has a type $\tau_k$ with $(k,\tau_k) \sim_\succeq (i,\tau')$, $(i,\tau)$ will be strictly happier if we move the cut to just below $(i,\tau)$. Therefore, it is still the case that every (bidder, type) pair who is made unhappier by moving the cut to just below $(i,\tau)$ is made strictly happier by moving the cut to just above $(i,\tau')$. So we can carry over the same reasoning as above (choosing $\epsilon, \delta$ sufficiently small), and again contradict the choice of $M^\succ$.

\end{itemize}
Therefore, it can not be the case that $i = j$.

\item \textbf{Case 2:} $i \neq j$ and there is never a cut below $(j,\tau')$.

This case is easy. If we switch $(j,\tau')$ and $(i,\tau)$ in $\succ$, then the set $S_\succ$ is exactly the same, and the distribution $M^\succ$ is exactly the same. However, we have now relabeled the types in $S_\succ$ to get closer to the top $|S_\succ|$ elements being in $S_\succ$. Note that all $\succeq$ with no cut below $(j,\tau')$ are all $\succ$-ordered for the new $\succ$ as well, so this is a valid swap.

\item \textbf{Case 3:} $i \neq j$ and there is sometimes a cut below $(j,\tau')$.

Pick a $\succeq$ in the support of $M^{\succ}$ that has a cut between $(j,\tau')$ and $(i,\tau)$ and in which $(j,\tau')$ gets the item with positive probability. Note that if such a $\succeq$ doesn't exist, we can remove the cut below $(j,\tau')$ in all $\succeq$ in the support of $M^\succ$ without changing the allocation probabilities and return to Case 2. From here, there are again two subcases:

\begin{itemize}
\item $(k,\tau'') \notin S_{\succ}$ for all $(k, \tau'') \sim_{\succeq} (j,\tau')$. This means that \emph{all} (bidder, type) pairs equivalent to $(j,\tau')$ are \emph{not} maximally unhappy. Therefore, if we pick a tiny $\epsilon$ and remove the cut below $(j,\tau')$ with probability $\epsilon$, only the types equivalent to $(j,\tau')$ will become unhappier. So there is a sufficiently small $\epsilon > 0$ for which this operation does not create any new maximally unhappy types. At the same time, because $i \neq j$ and $(j,\tau')$ receives the item with non-zero probability under $\succeq$, this operation makes $(i,\tau)$ strictly happier, as she now sometimes shares the item with $(j,\tau')$ (whereas previously she always lost). So this operation will create no new maximally unhappy (bidder, type) pairs, while making $(i,\tau)$ strictly happier, decreasing the size of $|S_\succ|$ and contradicting the choice of $M^\succ$.

\item There exists a $(k,\tau'') \in S_{\succ}$ with $(k,\tau'') \sim_{\succeq} (j, \tau')$ (and $(k,\tau'') \neq (j,\tau')$). Let $(k,\tau'')$ be the minimal such (bidder, type) pair under $\succ$. Note that by our choice of $(i,\tau)$, that all $(\ell, \tau''') \succ (k,\tau'')$ are also in $S_\succ$ (maximally unhappy). Now consider introducing a cut below $(k,\tau'')$ with some tiny probability $\epsilon$. Then the only (bidder, type) pairs who may become unhappier with this change are those that are still equivalent to $(j,\tau')$, and all such types are \emph{not} maximally unhappy. The only (bidder, type) pairs who may become happier with this change are those that are $(k,\tau'')$ or those that are $\succ (k,\tau'')$, all of which are in $S_\succ$. So if \emph{any} of these types become happier at all with this change, there is a sufficiently small probability $\epsilon$ with which we can make this change without introducing any new maximally unhappy types and therefore decreasing $|S_\succ|$, a contradiction. So we must not make any (bidder, type) pairs happier with this change, and therefore we must also not make any (bidder, type) pairs unhappier (note that it is impossible to make any (bidder, type) pair unhappier without making some other (bidder, type) pair happier, since we are treating $(0,\bot)$ as a regular type). So we may in fact introduce a cut below $(k,\tau'')$ with probability $1$ whenever $M^\succ$ samples $\succeq$ without affecting $\pi^{M^\succ}$ at all, but removing the original $\succeq$ from the support of $M^\succ$, and replacing it with a $\succeq$ in which all (bidder, type) pairs equivalent to $(j, \tau')$ are \emph{not} maximally unhappy. After doing this for all such $\succeq$, we must return to the previous sub-case, after which we again obtain a contradiction. 
\end{itemize}

{Hence, it can not be the case that  $i \neq j$ with a cut sometimes below $(j,\tau')$.}
\end{itemize}

At the end of all three cases, we see that if we ever have $(j,\tau') \notin S_\succ$ and $(i,\tau) \in S_\succ$, with $(j,\tau') \succ (i,\tau)$ and $(j,\tau'),(i,\tau')$ adjacent, then we must have $i \neq j$, and {no $\succeq$ in the support of $M^{\succ}$ ever places a cut directly below $(j,\tau')$}. Hence, we can simply swap the order of these types in $\succ$ without affecting $S_\succ$ or $F_\succ(M)$ (as we described in Case 2 above), and we do that repeatedly until $S_\succ$ is equal to the top $|S_\succ|$ (bidder, type) pairs according to $\succ$.

Now that we have shown the existence of such a $\succ$, we show that it implies a constricting set. Label the elements in $S_\succ$ as $(i_1,\tau_1) \succ \ldots \succ (i_k, \tau_k)$ ($k = |S_\succ|$). {Now consider a $\succeq$ in the support of $M^{\succ}$ that has no cut below $(i_k,\tau_k)$,} and consider putting a cut there with some tiny probability $\epsilon$ whenever $\succeq$ is sampled. The only effect this {\em might} have is that when the item is awarded to a (bidder, type) pair outside $S_\succ$, it is now awarded to a (bidder, type) pair inside $S_\succ$ instead with some probability. Therefore, if anyone gets happier, it is someone in $S_\succ$. However, if we make anyone in $S_\succ$ happier and choose $\epsilon$ small enough so that we don't make anyone outside of $S_\succ$ maximally unhappy, we decrease $|S_\succ|$, contradicting the choice of $M^\succ$. Therefore, putting a cut below $(i_k,\tau_k)$ cannot possibly make anyone happier, and therefore cannot make anyone unhappier. So we may w.l.o.g. assume that {there is a cut below $(i_k,\tau_k)$ in all $\succeq$ in the support of $M^{\succ}$}. But now we get that the item always goes to someone in $S_\succ$ {whenever a (bidder, type) pair in $S_\succ$ is reported}, yet all (bidder, type) pairs in this set are unhappy. Therefore, $S_{\succ}$ is a constricting set, {certifying that the given $\cR$ is infeasible}. 

Putting everything together, we have shown that if there is no $\succ$ with $F_\succ(M^\succ) = 0$ then the reduced form is infeasible. So there must be some $\succ$ with $F_\succ(M^\succ) = 0$, and such an $M^\succ$ induces the reduced form by sampling only $\succ$-ordered hierarchical allocation rules, completing the proof.
\end{prevproof}

\section{Multi-Item Mechanism Design} \label{sec:characterization}
In this section, we show how our results above on reduced forms can be useful for multi-item mechanism design as well. Essentially, our key observation is that an $m$-item interim allocation rule is feasible if and only if the $m$ projected single-item reduced forms are feasible, so the question is simply whether or not an $m$-item reduced form contains enough useful information for mechanism design. When buyers are additive, this information indeed suffices to guarantee that a mechanism inducing to this interim allocation rule is \emph{Bayesian Incentive Compatible}, which allows us to formulate and solve an optimization problem. We make these statements more precise shortly, but first provide some notation specific to multi-item mechanism design not covered in Section~\ref{sec:notation}.
\subsection{Notation}
For Section~\ref{sec:characterization}, there are $n$ bidders and $m$ items. All bidders' valuation functions are additive. We write $\vec{v}_i$ to denote the type of bidder $i$, with the convention that $v_{ij}$ represents her value for item $j$ and that her value for a bundle $S$ of items is simply $\sum_{j \in S} v_{ij}$. We still let $T$ denote the space of possible types, which is now a subset of $\mathbb{R}^n$.

To fully specify a (direct-revelation) multi-item mechanism for additive bidders, we need to describe, potentially succinctly, for all type profiles $\vec{v} \in T^n$, and for every bidder $i$, the outcome $M_i(\vec{v}) = (\vec{\phi}_i(\vec{v}), p_i({\vec{v}}))$ given by $M$ to bidder $i$ when the reported bidder types are $\vec{v}$. Here, $\phi_{ij}(\vec{v})$ is the ex-post probability that item $j$ is given to bidder $i$ when the reported types are $\vec{v}$, and $p_i({\vec{v}})$ is the ex-post price that $i$ pays. The {\em value} of bidder $i$ for outcome $M_i(\vec{w})$ is just her expected value $\vec{v}_i \cdot \vec{\phi}_i(\vec{w})$, {while her utility is {\em quasi-linear,} meaning that bidder $i$'s utility for the same outcome is $U(\vec{v}_i,M_i(\vec{w})):=\vec{v}_i \cdot \vec{\phi}_i(\vec{w}) - p_i(\vec{w})$.} The relation between the ex-post probabilities $\phi$ and interim probabilities $\pi$ is just the following: for all $i$,$j$, $\vec{v}_i \in T$: $\pi_{ij}(\vec{v}_i)=\mathbb{E}_{\vec{v}_{-i}\sim {\cal D}_{-i}}[\phi_{ij}(\vec{v}_i~;~\vec{v}_{-i})]$. We now formally define Bayesian Incentive Compatibility (BIC):

\begin{definition}(Bayesian Incentive Compatible Mechanism) A mechanism $M$ is called BIC iff the following inequality holds for all $i\in [n],\vec{v}_i,\vec{w}_i\in T$:
$$\mathbb{E}_{\vec{v}_{-i} \sim {\cal D}_{-i}}\left[U(\vec{v}_i,M_i(\vec{v}))\right] \ge \mathbb{E}_{\vec{v}_{-i} \sim {\cal D}_{-i}}\left[ U(\vec{v}_i,M_i(\vec{w}_i~;~\vec{v}_{-i})) \right].$$
\end{definition}

\subsection{Optimal Multi-Item Mechanism Design} \label{sec:optimal}
{We begin this section with our key observation, essentially stating that some single-item results (specifically, those in Section~\ref{sec:independent}) can be extended for free to some multi-item settings. Let us begin by being clear what we mean by an interim allocation rule projecting a reduced form onto item $j$.

\begin{definition}(Projected reduced form) Let there be $m$ heterogeneous items, $T$ be some arbitrary type space, and $\cR = \{\pi_{ij}(\cdot)\}_{i\in [n],j \in [m]}$ be some interim allocation rule of a mechanism. Then the projected reduced form of $\cR$ onto item $j$, $\cR_j$, is just $\cR_j = \{\pi_{ij}(\cdot)\}_{i \in [n]}$. Note that $\cR_j$ still takes as input types in the original type space $T$.
\end{definition}

\begin{observation}\label{obs:multi} An $m$-item interim allocation rule $\cR$ is feasible if and only if for all $j$, the projected reduced form $\cR_j$ onto item $j$ is feasible. Furthermore, if $\cR$ is feasible, $\cR$ is induced by the ex-post allocation rule that allocates each item $j$ according to $\cR_j$ independently of the others.
\end{observation}

\begin{proof}
First, assume that $\cR$ is feasible, and let $M$ be an ex-post allocation rule that induces $\cR$. We wish to come up with an allocation rule for item $j$ that induces $\cR_j$. Define $M_j${\footnote{{Note that $M_{j}$ takes the \emph{whole type} $\vec{v_{i}}$ of each bidder $i$ as input, and not just $v_{ij}$.}}} to be the {single-item} allocation rule that runs $M$ and awards the single item to whichever bidder was awarded item $j$ under $M$. Clearly, $M_j$ implements $\cR_j$, so if $\cR$ is feasible, so is each projection $\cR_j$.

Next, assume that each $\cR_j$ is feasible, and let $M_j$ be an ex-post allocation rule that induces $\cR_j$. Then let $M$ be the allocation rule that on every input type profile, runs $M_j$ on that type profile for all $j$ and awards item $j$ to whoever receives the single item under $M_j$. Clearly, the projection of the reduced form of $M$ onto item $j$ will be exactly $\cR_j$, so the reduced form of $M$ is exactly $\cR$. Therefore, $M$ induces $\cR$, and $\cR$ is feasible.

So $\cR$ is feasible if and only if each projection $\cR_j$ is feasible. Furthermore, the above argument shows that when $\cR$ is feasible, $\cR$ can be induced by an ex-post allocation rule that allocates each item separately.
\end{proof}}

Observation~\ref{obs:multi} combined with Theorem~\ref{thm:non-iidhierarchical} immediately yields our characterization of feasible multi-item interim allocation rules (from Section~\ref{sec:intro}). Replacing Theorem~\ref{thm:non-iidhierarchical} with Theorem~\ref{thm:iidhierarchical} provides a tighter characterization in the symmetric case. Note that at this point we have absolutely not addressed the issue of when this characterization is useful for multi-item mechanism design - all we have done is observed (somewhat trivially) that the structure of reduced forms is preserved under concatenation.

So now, let's address this issue and discuss the multi-item settings in which Observation~\ref{obs:multi} is useful for mechanism design. Essentially, we observe that the interim allocation rule as we have defined it provides sufficient information to determine whether or not a mechanism is BIC if and only if bidders' valuations are additive. We first present an example illustrating that this fails, for instance, when bidders are instead unit-demand.\footnote{A bidder is unit-demand if whenever they have value $v_j$ for item $j$, their value for set $S$ is equal to $\max_{j \in S} \{v_j\}$.}

\begin{example}\label{ex:interim} Consider a setting with a single unit-demand bidder, two items, and one possible type, $(1,1)$ (value one for each item). Consider the following two ex-post allocation rules:
\begin{itemize}
\item Pick $j$ uniformly random from $\{1,2\}$ and award item $j$.
\item Award the set of items $\{1,2\}$ with probability $1/2$, and the set $\emptyset$ with probability $1/2$.
\end{itemize}
Then these two ex-post allocation rules have the same interim allocation rule: $\pi_{11}(1,1) = \pi_{12}(1,1) = 1/2$. But the bidder's expected value under the first ex-post allocation rule is $1$, whereas under the second it is $1/2$. Therefore, the interim allocation rule simply does not contain enough information for the bidder to compute her expected value for reporting a given type to an ex-post allocation rule inducing it - because it depends on which ex-post allocation rule is chosen.

Note that if instead the bidder were additive, she would have expected value $1$ under both ex-post allocation rules, and this would not be an issue.
\end{example}

\begin{observation}\label{obs:additive}
When bidders are additive, the per-item interim allocation rule $\cR$ contains enough information for a bidder to compute her expected value when reporting type $\vec{w}_i$ to \emph{any} ex-post allocation rule inducing $\cR$. It is exactly $\sum_j v_{ij} \cdot \pi_{ij}(\vec{w}_i)$. 
\end{observation}
\begin{proof}
Let $M$ be any ex-post allocation rule with ex-post allocation probabilities $\phi_{ij}(\cdot)$. Then we can write the expected value of a buyer with type $\vec{v}_i$ for reporting $\vec{w}_i$ to $M$ as:
\begin{align*}
\mathbb{E}_{\vec{v}_{-i}\sim {\cal D}_{-i}}[\sum_j v_{ij} \cdot \phi_{ij}(\vec{w}_i~;~\vec{v}_{-i})]
 &= \sum_j \mathbb{E}_{\vec{v}_{-i}\sim {\cal D}_{-i}}[v_{ij} \cdot \phi_{ij}(\vec{w}_i~;~\vec{v}_{-i})]\\
& = \sum_j v_{ij} \cdot \mathbb{E}_{\vec{v}_{-i}\sim {\cal D}_{-i}}[\phi_{ij}(\vec{w}_i~;~\vec{v}_{-i})] = \sum_j v_{ij} \cdot \pi_{ij}(\vec{w}_i).
\end{align*}
\end{proof}

Essentially what makes additive buyers unique in comparison to other multi-dimensional valuation functions is that the marginal value for item $j$ is completely independent of the set it is being added to. Let us again emphasize that an ex-post allocation rule $M$ implementing an interim allocation rule $\cR$ absolutely takes into consideration the \emph{value} of bidders for items $\ell \neq j$ when determining how to allocate item $j$, as this information is stored in their types. However, Observations~\ref{obs:multi} shows that it need not consider how items $\ell \neq j$ are themselves \emph{allocated} when determining how to allocate item $j$. It is well-known that even when there is just a single additive bidder and two items, and even if the bidder's values for these items are distributed i.i.d., that the allocation rule of the revenue-optimal mechanism necessarily considers values for items $\ell \neq j$ when deciding the allocation of item $j$. Below is a folklore example {(that appears concretely, for instance, in~\cite{DaskalakisDT14})}. This shows that even if we are willing to restrict to a characterization only of interim allocation rules that are revenue-optimal for simple multi-item instances, we should not hope for a stronger characterization that (say) allocates each item $j$ independent of bidders' values for items $\ell \neq j$. 
\begin{example} There is a single additive bidder and two items. Each $v_j$ is drawn independently and uniformly from the set $\{1,2\}$. Then the revenue optimal mechanism awards both items to the bidder whenever $v_1 + v_2 \geq 3$, and charges price $3$. 

Note that when $v_1 = 1$, whether or not the bidder receives item $1$ depends on $v_2$ (namely, she receives item $1$ iff $v_2 = 2$). 
\end{example}

With all this in mind, let us now formally state the multi-item auction problem we solve:

\smallskip\noindent  \framebox{
\begin{minipage}{.975\hsize}
{\bf BIC Multi-item auction.} Given as input {$n$} distributions $\mathcal{D}_1,\ldots,\mathcal{D}_n$ over valuation vectors for $m$ items, output a BIC mechanism $M$ whose expected revenue is optimal relative to any other, possibly randomized, BIC mechanism, when played by $n$ additive bidders whose valuation vectors are sampled independently from $\mathcal{D}_1,\ldots,\mathcal{D}_n$. Note that each $\mathcal{D}_i$ need not be a product distribution, a single bidder's values for different items may be arbitrarily correlated.
\end{minipage}}

\smallskip {Our approach to solving proving Theorem~\ref{thm:computation main} is to use the separation oracle {for checking the feasibility of a reduced form developed in} Corollary~\ref{cor:grand theorem independent} inside a {linear program that optimizes over all feasible interim allocation rules and interim price rules}. }{We begin with the LP formulation in Figure~\ref{fig:LP} below, followed by a proof that the LP is correct.

\begin{figure}[h]
\colorbox{MyGray}{
\begin{minipage}{\textwidth} {
\noindent\textbf{Variables:}
\begin{itemize}
\item $\pi_{ij}(\vec{v}_i)$, for all bidders $i\in[n]$, items $j\in[m]$, and $\vec{v}_i \in T$, the interim probability that bidder $i$ gets item $j$ when reporting type $\vec{v}_i$ ({$mnc$ variables}).
\item $q_i(\vec{v}_i)$, for all bidders $i\in[n]$, $\vec{v}_i \in T$ , the interim expected price that bidder $i$ pays when reporting type $\vec{v}_i$ ({$nc$ variables}).
\end{itemize}
\textbf{Constraints:}
\begin{itemize}
\item $0 \leq \pi_{ij}(\vec{v}_i) \leq 1$, for all $i\in[n]$, $j\in[m]$, $\vec{v}_i \in T$, guaranteeing that each $\pi_{ij}(\vec{v}_i)$ is a probability ({$mnc$ constraints}).
\item $\sum_{j\in[m]} v_{ij}\pi_{ij}(\vec{v}_i) - q_i(\vec{v}_i) \geq 0$, for all $i\in[n]$, $\vec{v}_i \in T$, guaranteeing that the mechanism is interim Individually Rational (interim IR) ({$nc$ constraints}).
\item $\sum_{j\in[m]} v_{ij}\pi_{ij}(\vec{v}_i) - q_i(\vec{v}_i) \geq \sum_{j\in[m]} v_{ij}\pi_{ij}(\vec{v}'_i) - q_i(\vec{v}'_i)$, for all $i\in [n],\vec{v}_i,\vec{v}'_i \in T$, guaranteeing that the mechanism is Bayesian Incentive Compatible (BIC) ({$nc^{2}$ constraints}).
\item $SO(\vec{\pi},\mathcal{D}) = $``\textsc{Yes}'', guaranteeing that there is an ex-post allocation rule inducing $\pi$;
\end{itemize}
\textbf{Maximizing:}
\begin{itemize}
\item $\sum_{i\in[n],\vec{v}_i \in T} q_i(\vec{v}_i)\Pr[\vec{v}_i \leftarrow \mathcal{D}_i]$, the expected revenue.
\end{itemize}
}
\end{minipage}} \caption{A folklore LP (that appears concretely, e.g., in~\cite{DaskalakisW12}), where we use a separation oracle to determine feasibility of the interim allocation rule. In parentheses at the end of each line is the number of such variables/constraints .}\label{fig:LP}
\end{figure}

\begin{proposition}\label{prop:LP} Provided that $SO$ acts as a valid separation oracle for the space of feasible interim allocation rules, the LP of Figure~\ref{fig:LP} outputs the revenue optimal interim allocation rule and interim price rule for BIC multi-item auction in time polynomial in $n,m,c$ and the runtime of $SO$.
\end{proposition}

\begin{proof}
First, it is clear that any output of the LP of Figure~\ref{fig:LP} (henceforth, just ``the LP'') is interim IR, BIC, and feasible, as long as $SO$ is correct. This is because the constraints on an interim allocation rule/price rule pair to be interim IR are linear and explicitly included in the LP. The same holds for BIC. Therefore, as long as $SO$ is correct, any interim allocation rule/price rule pair accepted by the LP must be interim IR, BIC, and feasible. It is also clear that the objective function correctly computes the expected revenue of any interim price rule considered. So the LP outputs exactly the feasible, interim IR, BIC interim allocation rule/price rule pair that maximizes expected revenue with respect to all feasible, interim IR, BIC interim allocation rule/price rule pairs.

Second, it is clear that every feasible, IR, BIC mechanism has an interim allocation rule/price rule pair, and that this pair is also interim IR, feasible, and BIC. So the revenue-optimal interim allocation/price rule pair output by the LP is indeed optimal with respect to all feasible, IR, BIC mechanisms.

Finally, it is clear that the number of variables and constraints (excluding $SO$) is polynomial in $nmc$. Therefore, the LP can be solved in time polynomial in $n,m,c$ and the runtime of $SO$ via a direct application of Theorem~\ref{thm:GLS}.
\end{proof}

\begin{prevproof}{Theorem}{thm:computation main}
{Theorem~\ref{thm:computation main} now follows immediately from Proposition~\ref{prop:LP}, Observation~\ref{obs:multi}, and Corollary~\ref{cor:grand theorem independent}. Corollary~\ref{cor:grand theorem independent} combined with Observation~\ref{obs:multi} guarantees that we can design the desired separation oracle, which terminates in time $\poly(n, m, c)$, and Proposition~\ref{prop:LP} guarantees that we can use this to solve the LP in time $\poly(n, m, c)$. {After finding the optimal interim allocation/price rule pair $(\vec{\pi}^*,\vec{q}^*)$, Corollary~\ref{cor:grand theorem independent} shows how to find in time $\poly(n, m, c)$ a succinct description of an ex-post allocation rule implementing $\vec{\pi}^*$ (as a distribution over at most $cn+1$ strict hierarchical allocation rules). So one can implement the optimal mechanism by using this ex-post allocation rule, and charging bidder $i$ price $q_i^*(\vec{v}_i)$ when her bid is $\vec{v}_i$}.}
\end{prevproof}

We also note that the LP of Figure~\ref{fig:LP} only requires that the mechanism be interim individually rational. A well-known simple trick (shown e.g. in~\cite{DaskalakisW12}) converts any BIC, interim IR mechanism into one that is BIC and ex-post IR with no loss in revenue, so this is w.l.o.g.} For completeness, we describe the trick below. Note that the trick does not require any assumptions on the valuation functions of bidders except that they are quasi-linear and risk-neutral.

\begin{proposition}\label{prop:trick}
Let $M$ be any BIC and interim IR mechanism for quasi-linear and risk-neutral bidders. There exists another mechanism $M'$ that uses exactly the same ex-post allocation rule and interim price rule as $M$ that is also ex-post IR. 
\end{proposition}
\begin{proof}
For any bidder $i$ and valuation function/type $v_i(\cdot): 2^{[m]} \rightarrow \mathbb{R}$ ($v_i$ takes as input a set of items and outputs a value), let $V^M_i(v_i)$ denote the expected value of bidder $i$ when truthfully reporting type $v_i$ to $M$ (over any randomness in $M$, and the randomness in other bidders' types). Let also $q^M_i(v_i)$ denote the interim price paid by $i$ when reporting type $v_i$ to $M$ (over the same randomness). As $M$ is interim IR, we must have $q^M_i(v_i) \leq V^M_i(v_i)$. 

Now define the following ex-post price rule for $M'$: whenever bidder $i$ receives items $S$ after reporting type $v_i$, charge her $\frac{q^M_i(v_i)}{V^M_i(v_i)} \cdot v_i(S)$. Observe that we have not changed the ex-post allocation rule, so clearly $M$ and $M'$ have the same ex-post allocation rule. As $\frac{q^M_i(v_i)}{V^M_i(v_i)} \leq 1$, $M'$ is clearly ex-post IR. All that remains to verify is that $M$ and $M'$ have the same interim price rule. But this is also clear: if we define $\phi_S(\vec{v})$ to be the probability that $M$ allocates set $S$ to bidder $i$ on input $\vec{v}$, then the interim price paid by bidder $i$ when reporting type $v_i$ to $M'$ is exactly:
\begin{align*}
\mathbb{E}_{\vec{v}_{-i}\sim {\cal D}_{-i}}[\sum_{S \subseteq [m]}\phi_S(\vec{v}) \cdot \frac{q^M_i(v_i)}{V^M_i(v_i)} \cdot v_i(S)] &=\frac{q^M_i(v_i)}{V^M_i(v_i)} \cdot \mathbb{E}_{\vec{v}_{-i}\sim {\cal D}_{-i}}[\sum_{S \subseteq [m]}\phi_S(\vec{v}) \cdot v_i(S)]\\
&= \frac{q^M_i(v_i)}{V^M_i(v_i)} \cdot V^M_i(v_i) = q^M_i(v_i).
\end{align*}

\end{proof}

\section{Conclusions and Discussion}
Motivated by settings where an optimization approach is necessary to develop optimal auctions, we study single-item reduced-form auctions for asymmetric bidders. We provide a linear-sized subset of necessary and sufficient Border conditions (down from exponential-sized of prior work) that can be checked in nearly-linear time, and also show that every feasible reduced form can be implemented as a distribution of hierarchical allocation rules that all respect the same total ordering over all bidders' types. We further show that our results imply both polynomial-time algorithms and structural characterizations for multi-item auctions with additive bidders. 

Our work demonstrates how a better understanding of reduced-form auctions in the core single-item setting can be utilized for much more general settings, and also that a computational lens can lead not only to tractable optimization algorithms, but also improved structural understanding. 

We conclude with a very brief discussion on how one could extend our approach for multi-item auction design beyond additive buyers. Indeed, the core issue is that the interim allocation probabilities no longer contain enough information to verify that a mechanism is BIC (Example~\ref{ex:interim}), so in order to get mileage out of our approach one first needs to come up with a new interim description of auctions. It is not hard to come up with definitions that work for, say, unit-demand buyers: first observe that it is w.l.o.g. to only consider auctions which award each bidder at most one item.\footnote{To see this, take any auction that awards multiple items to bidder $i$ with reported type $\vec{v}_i$, and modify it to throw away all items except the highest value item (according to $\vec{v}_i$). Then if bidder $i$ was telling the truth, their utility doesn't change at all. If she was lying, her utility might go down. So the mechanism is still truthful, and the revenue is the same because the payments are the same.} Then simply define $\pi_{ij}(\vec{v}_i)$ to be the interim probablity that bidder $i$ receives item $j$ when submitting valuation vector $\vec{v}_i$, and exactly the same LP as written in Figure~\ref{fig:LP} finds the revenue-optimal BIC mechanism, provided we can design a computationally efficient separation oracle for this new space of feasible interim allocation rules (which never allocate the same item to multiple bidders \emph{or} the same bidder multiple items). However, obtaining such a separation oracle is no longer simply the product of $n$ single-item problems, and in fact Gopalan et. al. prove that, under well-believed complexity-theoretic assumptions, no computationally efficient separation oracle for this space exists~\cite{GopalanNR15}. 

Still, our approach has proven useful for the design of \emph{nearly}-optimal multi-item auctions in settings with unit-demand buyers (and in fact significantly more general settings as well) in follow-up works by the authors and others~\cite{CaiDW12b, CaiDW13a, CaiDW13b, BhalgatGM13, DaskalakisDW15, CaiDW16}, essentially by making use of computationally-efficient but \emph{approximate} characterization of the feasible interim allocation probabilities\notshow{Border-like theorems} in these settings. The present paper remains unique in this line of works for finding the optimal mechanism without approximation error.

\bibliographystyle{alpha}
\bibliography{masterBib}
\appendix
\section{Omitted Proofs}\label{app:proofs}

The first omitted proof is of Corollary~\ref{cor:algiid}, which simply claimed that we can find a violated Border constraint in the symmetric case in time $O(c(\log c + \log n))$. 

\begin{proof}[Proof of Corollary~\ref{cor:algiid}]
First, sort the types $\tau$ in $T$ in decreasing order according to $\pi(\tau)$ in time $O(c \log c)$, and label the distinct values of $\pi(\tau)$ as $x_1,\ldots, x_k$ ($k \leq c$), and define $X_i = \{\tau~|~\pi(\tau) = x_i\}$. Next, compute $\sum_{\tau: \pi(\tau) \geq x_i} \pi(\tau) \Pr[\tau]$ for all $i$ in the following way: First, compute $\sum_{\tau \in X_i} \pi(\tau) \Pr[\tau]$ for all $i$. This can clearly be done in time $O(|X_i|)$ for all $i$, and therefore the total computation takes time $O(c)$. Next, observe that $\sum_{\tau:\pi(\tau) \geq x_{i+1}} \pi(\tau) \Pr[\tau] = \sum_{\tau:\pi(\tau) \geq x_{i}} \pi(\tau) \Pr[\tau] + \sum_{\tau \in X_{i+1}} \pi(\tau) \Pr[\tau]$. So we can compute $\sum_{\tau:\pi(\tau) \geq x_{i+1}} \pi(\tau) \Pr[\tau]$ from $\sum_{\tau:\pi(\tau) \geq x_{i}} \pi(\tau) \Pr[\tau]$ using just $O(1)$ additional computation. Therefore, we can compute $\sum_{\tau: \pi(\tau) \geq x_i} \pi(\tau) \Pr[\tau]$ for all $i$ in total time $O(c)$. This immediately gives us the LHS for all inequalities~\eqref{eq:border's condition} in total time $O(c\log c)$. Similarly, we can compute $\sum_{\tau \in X_i} \Pr[\tau]$ for all $i$ in total time $O(c)$, and use these in the same way to compute $\Pr_{t \sim D_1}[\pi(t)\geq x_i] = \sum_{\tau: \pi(\tau) \geq x_i} \Pr[\tau]$ for all $i$ in total time $O(c\log c)$. With this, we can then compute the RHS of each inequality~\eqref{eq:border's condition} using repeated squaring in time $O(\log n)$ per inequality. 

So all together, we have a pre-processing stage which takes time $O(c \log c)$, and an additional $O(c\log n)$ to compute each RHS, resulting in a total runtime of $O(c(\log c + \log n))$. 
\end{proof}

The next omitted proof is of (the algorithmic portion of) Theorem~\ref{thm: independent}, which claims that we can find a violated Border constraint in the asymmetric case in time $O(cn\cdot \log(cn))$. 

\begin{prevproof}{Theorem}{thm: independent}
{ Proposition~\ref{prop: virtualsuffices} immediately yields the first part of Theorem~\ref{thm: independent} (that a reduced form is feasible if and only if it satisfies Equation~\ref{eq:our virtual feasibility condition} for all $x$). The only remaining detail to check is that a violated Border constraint (if it exists) can be found in the desired runtime, which follows from a similar routine computation as in Corollary~\ref{cor:algiid}. For the sake of completeness, we include with a full proof of this. 

First, sort all (bidder,  type) pairs $(i, \tau) \in [n] \times T$ in decreasing order according to $\hat{\pi}_i(\tau)$ in time $O(nc\log(nc))$ ($c=|T|$), label the distinct values of $\hat{\pi}_i(\tau)$ as $x_{1},\ldots, x_{k}$ {($k \leq nc$)}, and store $X_{j} = \{(i, \tau)\ |\ \hat{\pi}_i(\tau) = x_{j}\}$ (i.e. $X_j$ is the set of (bidder, type) pairs for which $\hat{\pi}_i(\tau) = x_j$). Next, to compute all left-hand sides of Equation~\eqref{eq:our virtual feasibility condition}, begin by computing $\sum_{(i,\tau) \in X_j} \Pr[t_i = \tau]\cdot \pi_i(\tau)$ for all $j$. This can clearly be done in time $O(|X_j|)$ for all $j$, and therefore the total computation takes time $O(nc)$. Next, observe that $\sum_{(i,\tau):\hat{\pi}_i(\tau) \geq x_{j+1}} \Pr[t_i = \tau]\cdot \pi_i(\tau) = \sum_{(i,\tau):\hat{\pi}_i(\tau) \geq x_j} \Pr[t_i = \tau]\cdot \pi_i(\tau) + \sum_{(i,\tau) \in X_{j+1}} \Pr[t_i = \tau] \cdot \pi_i(\tau)$. So we can compute $\sum_{(i,\tau):\hat{\pi}_i(\tau) \geq x_{j+1}} \Pr[t_i = \tau]\cdot \pi_i(\tau)$ from $\sum_{(i,\tau):\hat{\pi}_i(\tau) \geq x_j} \Pr[t_i = \tau]\cdot \pi_i(\tau)$ with $O(1)$ additional computation (by making use of our pre-processed values). So starting with $x_1$, we can compute all left-hand sides of Equation~\eqref{eq:our virtual feasibility condition} with $O(1)$ additional computation per constraint, for a total computation time of $O(nc)$. }

{Computing all right-hand sides is a touch trickier, because we don't actually want to multiply $n$ numbers together for each of $nc$ inequalities. So we will make use of the fact that most of the terms going into the product on the RHS don't change between successive inequalities. To compute all right-hand sides, additionally for all $i$, sort all types $\tau \in T$ in decreasing order of $\hat{\pi}_i(\tau)$ in time $O(c \log c)$ per bidder, or $O(nc \log c)$ in total, label the distinct values of $\hat{\pi}_i(\tau)$ as $x_{i1},\ldots, x_{ik_i}$ ($k_i \leq c$), and store $X_{ij} = \{\tau\ |\ \hat{\pi}_i(\tau) = x_{ij}\}$. Compute $\sum_{\tau \in X_{ij}} \Pr[t_i = \tau]$ for all $i, j$ in total time $O(nc)$. {Recall $S_x^{(i)}=\{\tau\ |\ \hat{\pi}_i(\tau)\geq x \}$, and observe that $\Pr_{t_i \sim \mathcal{D}_i}[t_i \in S_{x_{ij}}^{(i)}] = \sum_{\tau:\hat{\pi}_i(\tau)\geq x_{ij}} \Pr[t_i = \tau]$ can be computed for fixed $i$ and all $j$ in time $O(c)$, and therefore for all $i$ and all $j$ in total time $O(nc)$.} With this, finally compute $\frac{1-\Pr_{t_i \sim \mathcal{D}_i}[t_i \in S_{x_{i(j+1)}}^{(i)}]}{1-\Pr_{t_i \sim \mathcal{D}_i}[t_i \in S_{x_{ij}}^{(i)}]}$ for all $i$ and $j \in \{0,\ldots,k_i-1\}$ (denoting $S_{x_i0}^{(i)} = \emptyset$), again in time $O(c)$ per bidder, or $O(nc)$ in total (it is just at most one division per $(i, \tau) \in [n] \times T$). Now, starting from $j=1$, we will inductively compute $\prod_i (1-\Pr_{t_i \sim \mathcal{D}_i}[t_i \in S_{x_j}^{(i)}])$ for all $j$, using the result for $j-1$ as well as our precomputed terms. Observe that $\prod_i (1-\Pr_{t_i \sim \mathcal{D}_i}[t_i \in S_{x_j}^{(i)}]) = \prod_i (1-\Pr_{t_i \sim \mathcal{D}_i}[t_i \in S_{x_{j-1}}^{(i)}]) \cdot \prod_i \frac{1-\Pr_{t_i \sim \mathcal{D}_i}[t_i \in S_{x_j}^{(i)}]}{1-\Pr_{t_i \sim \mathcal{D}_i}[t_i \in S_{x_{j-1}}^{(i)}]}$. For many $i$, it might be that $\frac{1-\Pr_{t_i \sim \mathcal{D}_i}[t_i \in S_{x_j}^{(i)}]}{1-\Pr_{t_i \sim \mathcal{D}_i}[t_i \in S_{x_{j-1}}^{(i)}]} = 1$ (namely, all $i$ such that $(i,\tau) \notin X_j$ for all $\tau \in T$). So to compute $\prod_i \frac{1-\Pr_{t_i \sim \mathcal{D}_i}[t_i \in S_{x_j}^{(i)}]}{1-\Pr_{t_i \sim \mathcal{D}_i}[t_i \in S_{x_{j-1}}^{(i)}]}$ simply multiply together for all $i$ such that $\exists \tau \in T, (i,\tau) \in X_j$ the corresponding $\frac{1-\Pr_{t_i \sim \mathcal{D}_i}[t_i \in S_{x_j}^{(i)}]}{1-\Pr_{t_i \sim \mathcal{D}_i}[t_i \in S_{x_{j-1}}^{(i)}]}$, which we have precomputed. This can clearly be done in time $O(|X_j|)$. So the total time to compute $\prod_i (1-\Pr_{t_i \sim \mathcal{D}_i}[t_i \in S_{x_j}^{(i)}])$ for all $j$ is $O(\sum_j |X_j|) = O(nc)$. From here, we just take $1-\prod_i (1-\Pr_{t_i \sim \mathcal{D}_i}[t_i \in S_{x_j}^{(i)}])$ to get the corresponding right-hand sides for all $x_j$. So all inequalities of the form~\eqref{eq:our virtual feasibility condition} can be checked in total time $O(nc \log (nc))$. 
}
\end{prevproof}

The final omitted proof is a technical lemma used inside the proof of Theorem~\ref{thm: independentcorners}. The lemma states that monotonicity constraints can be tight (and linearly independent) only if they are between two types in the same equivalence class. 

\begin{prevproof}{Lemma}{lem:tightmono}
There are six cases to consider. Below we are studying which monotonicity constraints might potentially be tight in \emph{any} possible ex-post allocation rule that induces $\tilde{\pi}$. 
\begin{itemize}
\item What if $(i,\tau_{i,j+1}) \succeq (i,\tau_{i,j})$? Then $(i,j)^{th}$ monotonicity constraint might be tight.
\item What if $(i,\tau_{i,j}) \succeq (0,\bot) \succeq (i,\tau_{i,j+1})$? Then by definition of $\succeq$, $\tilde{\pi}_i(\tau_{i,j}) > 0 = \tilde{\pi}_{i}(\tau_{i,j+1})$, so the $(i,j)^{th}$ monotonicity constraint can't be tight.
\item What if $(0,\bot) \not \succeq (i,\tau_{i,j+1}) \not \succeq (i,\tau_{i,j})$ and there exists a bidder {$k\neq i$} and type $\tau'$ with $(i,\tau_{i,j+1}) \not \succeq (k,\tau') \not \succeq (i,\tau_{i,j})$? First, it is clear that bidder $i$ with type $\tau_{i,j+1}$ can \emph{only} win when $(\tau_{i,j+1}) \succeq (\ell, t_\ell)$ for all $\ell \neq i$ (because on any other profile, there exists a (bidder, type) pair present in a tight Border constraint without $(i,\tau_{i,j+1})$), and that bidder $i$ with type $\tau_{i,j}$ \emph{must} win in all these cases (exactly because on all such profiles, $(i,\tau_{i,j})$ is present in a tight Border constraint without $(\ell, t_\ell)$ for all $\ell$). So for every opposing profile where $\tau_{i,j+1}$ has a chance of winning, $\tau_{i,j}$ certainly wins. Now we show that an opposing profile exists where bidder $i$ loses with type $\tau_{i,j+1}$ but wins with type $\tau_{i,j}$, meaning that the $(i,j)^{th}$ monotonicity constraint can't be tight. Because $(i,\tau_{i,j+1}) \succeq (0,\bot)$, we necessarily have $(i,\tau_{i,j+1}) \succeq (\ell,\tau_{\ell,c_\ell})$ for all bidders $\ell$. So consider the opposing profile where bidder $k$ has type $\tau'$, and all other bidders $\ell\notin \{i,k\}$ have type $\tau_{\ell, c_\ell}${, which arises with non-zero probability by our choice of $c_\ell$}. Clearly, bidder $i$ with type $\tau_{i,j}$ must win the item against these opponents, while bidder $i$ with type $\tau_{i,j+1}$ must lose. Therefore, the $(i,j)^{th}$ monotonicity constraint can't be tight.

\item What if $(0,\bot) \not \succeq (i,\tau_{i,j+1}) \not \succeq (i,\tau_{i,j})$ and there exists a bidder $k\neq i$ and type $\tau'$ with $(i,\tau_{i,j}) \succeq (k,\tau') \succeq (i,\tau_{i,j})$? We know that bidder $i$ with type $\tau_{i,j+1}$ \emph{must} lose whenever there exists a bidder $\ell$ with $(\ell,\tau_\ell) \succeq (i,\tau_{i,j})$. However, if bidder $i$ also loses with type $\tau_{i,j}$ against \emph{all} such profiles, then the Border constraint for the set of all (type, bidder) pairs  $(\ell,\tau) \neq (i,\tau_{i,j})$ with $(\ell,\tau) \succeq (i,\tau_{i,j})$ would necessarily be tight as well, meaning that the type $(k,\tau')$ is present in a tight Border constraint that $(i,\tau_{i,j})$ is not, contradicting $(i,\tau_{i,j}) \succeq (k,\tau')$.\footnote{{Observe that all $(\ell,\tau)$ with $(\ell,\tau) \succeq(i,\tau_{i,j})$ are necessarily have $\tau \neq \tau_{\ell,c}$, as such types are present in at most one tight Border constraint (namely~\eqref{eq: better}), in which $(i,\tau_{i,j+1})$ is itself present. So the specified Border constraint is indeed one that remains in Equation~\eqref{eq: simpler}.}} So bidder $i$ with type $\tau_{i,j}$ must win with non-zero probability against \emph{some} such opposing profile, meaning again that the $(i,j)^{th}$ monotonicity constraint can't be tight.

\item What if $(0,\bot) \not \succeq (i,\tau_{i,j+1}) \not \succeq (i,\tau_{i,j})$ and there exists a bidder $k\neq i$ and type $\tau'$ with $(i,\tau_{i,j+1}) \succeq (k,\tau') \succeq (i, \tau_{i,j+1})$? The reasoning is symmetric to the above case. We know that bidder $i$ with type $\tau_{i,j}$ \emph{must} win whenever all other bidders $\ell$ have $(i,\tau_{i,j+1}) \succeq (\ell,t_\ell)$. However, if bidder $i$ also wins with type $\tau_{i,j+1}$ against \emph{all} such profiles, then the Border constraint for the set $(i,\tau_{i,j+1})$ and all (type, bidder) pairs $(\ell,\tau)$ with $(i,\tau_{i,j+1}) \not \succeq (\ell,\tau)$ {(i.e. the set $\{(\ell, \tau) | (i, \tau_{i, j+1}) \not \succeq (\ell, \tau)\} \cup \{(i, \tau_{i, j+1})\}$)}
would necessarily be tight as well, meaning that the type $(i,\tau_{i,j+1})$ is in a tight Border constraint that $(k,\tau')$ is not, contradicting $(k,\tau')\succeq (i,\tau_{i,j+1})$.\footnote{{Again, observe that the specified Border constraint is indeed one that remains in Equation~\eqref{eq: simpler}, because if $j+1=c$, then $\tilde{\pi}_{k}(\tau')$ will be $0$ contradicting the assumption that $(k,\tau') \succeq (i,\tau_{i,j+1}) \succeq (0,\bot)$.}} So bidder $i$ with type $\tau_{i,j+1}$ must lose with non-zero probability against \emph{some} such opposing profile, meaning again that the $(i,j)^{th}$ monotonicity constraint can't be tight.

\item What if $(0,\bot) \not \succeq (i,\tau_{i,j+1}) \not \succeq (i,\tau_{i,j})$ and for all bidders $k \neq i$ and types $\tau'$, $(i,\tau_{i,j}) \not \succeq (k,\tau')$ or $(k,\tau') \not \succeq (i,\tau_{i,j+1})$? In this case, the $(i,j)^{th}$ monotonicity constraint is guaranteed to be tight, but linearly dependent on the set of tight Border constraints. {We first argue that the Border constraint is tight for the set $A=\{(k,\tau') | (k,\tau') \succeq (i,\tau_{i,j})\} - (i,\tau_{i,j})$: on any profile $(t_1,\ldots,t_n)$ where $t_i \neq \tau_{i,j}$, but some $(k,t_k)\succeq (i,\tau_{i,j})$ (possibly $k = i$), clearly some bidder $k$ with $(k,t_k) \in A$ must win (as the Border constraint for $B = \{(k,\tau') | (k,\tau') \succeq (i,\tau_{i,j})\}$ is tight, by definition of $\succeq$). Moreover, when $t_i = \tau_{i,j}$ and the profile $(t_1,\ldots, t_n)$ has some type $(k,t_k)\in A$ then $k\neq i$, and we must have $(i,\tau_{i,j})\not\succeq (k,t_k)$ (by hypothesis defining this case). Therefore, the winner must still be some $(k,t_k) \in A$, and we conclude that the Border constraint for $A$ must be tight. 

Similar reasoning implies that the Border constraint for set $C = \{(k,\tau') | (k,\tau') \succeq (i,\tau_{i,j})\} \cup \{(i,\tau_{i,j+1})\}$ is tight.} Finally, one can take a linear combination of the tight Border constraints for the three sets $\{(k,\tau') | (k,\tau') \succeq (i,\tau_{i,j})\} - (i,\tau_{i,j})$, $\{(k,\tau') |(k,\tau') \succeq (i,\tau_{i,j})\}$ and $\{(k,\tau') | (k,\tau') \succeq (i,\tau_{i,j})\} \cup \{(i,\tau_{i,j+1})\}$ to recover the $(i,j)^{th}$ monotonicity constraint.\footnote{Again, all three Border constraints remain in Equation~\eqref{eq: simpler}.} To be thorough, we do the calculation below: we first list the three tight Border's constraints.

\begin{equation}\label{eq:Border set A}
	\sum_k \sum_{j | (k,\tau_{k,j}) \in A}\Pr[t_{k}=\tau_{k,j}] \tilde{\pi}_k(\tau_{k,j}) = 1- \prod_k \left(1-\sum_{j |(k,\tau_{k,j}) \in A} \Pr[t_k = \tau_{k,j}]\right) 
\end{equation}

\begin{equation}\label{eq:Border set B}
\sum_k \sum_{j | (k,\tau_{k,j}) \in B}\Pr[t_{k}=\tau_{k,j}] \tilde{\pi}_k(\tau_{k,j}) = 1- \prod_k \left(1-\sum_{j |(k,\tau_{k,j}) \in B} \Pr[t_k = \tau_{k,j}]\right) 
\end{equation}

\begin{equation}\label{eq:Border set C}
\sum_k \sum_{j | (k,\tau_{k,j}) \in C}\Pr[t_{k}=\tau_{k,j}] \tilde{\pi}_k(\tau_{k,j}) = 1- \prod_k \left(1-\sum_{j |(k,\tau_{k,j}) \in C} \Pr[t_k = \tau_{k,j}]\right)
\end{equation}
 
 Clearly, Equation~\eqref{eq:Border set A} and~\eqref{eq:Border set B} imply that 
 \begin{equation}\label{eq:set A and B}
 	\Pr[t_i = \tau_{i,j}]\cdot \tilde{\pi}_i(\tau_{i,j}) = \prod_{k \neq i}\left(1-\sum_{j | (k, \tau_{k,j})\in A} \Pr[t_k=\tau_{k,j}]\right) \cdot \Pr[t_i = \tau_{i,j}].
 \end{equation}
 
 Also, Equation~\eqref{eq:Border set B} and~\eqref{eq:Border set C} imply that 
 \begin{align}
 	\Pr[t_i = \tau_{i,j+1}]\cdot \tilde{\pi}_i(\tau_{i,j+1}) &= \prod_{k \neq i}\left(1-\sum_{j | (k, \tau_{k,j})\in B } \Pr[t_k=\tau_{k,j}]\right) \cdot \Pr[t_i = \tau_{i,j+1}]\nonumber \\
&=\prod_{k \neq i}\left(1-\sum_{j | (k, \tau_{k,j})\in A } \Pr[t_k=\tau_{k,j}]\right) \cdot \Pr[t_i = \tau_{i,j+1}].
 \end{align}
 
 Hence, 
$$ \tilde{\pi}_{i,j} (\tau_{i,j}) = \tilde{\pi}_{i,j+1} (\tau_{i,j+1})= \prod_{k \neq i}\left(1-\sum_{j | (k, \tau_{k,j})\in A } \Pr[t_k=\tau_{k,j}]\right).$$

\notshow{\begin{align*}
\sum_k \sum_{j | (k,\tau_{k,j}) \in A}\Pr[t_{k}=\tau_{k,j}] \tilde{\pi}_k(\tau_{k,j}) = &1- \prod_k \left(1-\sum_{j |(k,\tau_{k,j}) \in A} \Pr[t_k = \tau_{k,j}]\right) \\
\sum_k \sum_{j | (k,\tau_{k,j}) \in B}\Pr[t_{k}=\tau_{k,j}] \tilde{\pi}_k(\tau_{k,j}) = &1- \prod_k \left(1-\sum_{j |(k,\tau_{k,j}) \in B} \Pr[t_k = \tau_{k,j}]\right) \\
\Rightarrow \Pr[t_i = \tau_{i,j}]\cdot \tilde{\pi}_i(\tau_{i,j}) &= \prod_{k \neq i}\left(1-\sum_{j | (k, \tau_{k,j})\in A} \Pr[t_k=\tau_{k,j}]\right) \cdot \Pr[t_i = \tau_{i,j}].\\
\sum_k \sum_{j | (k,\tau_{k,j}) \in B}\Pr[t_{k}=\tau_{k,j}] \tilde{\pi}_k(\tau_{k,j}) = &1- \prod_k \left(1-\sum_{j |(k,\tau_{k,j}) \in B} \Pr[t_k = \tau_{k,j}]\right) \\
\sum_k \sum_{j | (k,\tau_{k,j}) \in C}\Pr[t_{k}=\tau_{k,j}] \tilde{\pi}_k(\tau_{k,j}) = &1- \prod_k \left(1-\sum_{j |(k,\tau_{k,j}) \in C} \Pr[t_k = \tau_{k,j}]\right) \\
\Rightarrow \Pr[t_i = \tau_{i,j+1}]\cdot \tilde{\pi}_i(\tau_{i,j+1}) &= \prod_{k \neq i}\left(1-\sum_{j | (k, \tau_{k,j})\in B } \Pr[t_k=\tau_{k,j}]\right) \cdot \Pr[t_i = \tau_{i,j+1}]\\
&=\prod_{k \neq i}\left(1-\sum_{j | (k, \tau_{k,j})\in A } \Pr[t_k=\tau_{k,j}]\right) \cdot \Pr[t_i = \tau_{i,j+1}].\\
\end{align*}

Hence, 
$$ \tilde{\pi}_{i,j} = \tilde{\pi}_{i,j+1} = \prod_{k \neq i}\left(1-\sum_{j | (k, \tau_{k,j})\in A } \Pr[t_k=\tau_{k,j}]\right).$$}
\end{itemize}
From this case analysis, we now conclude that the $(i,j)^{th}$ monotonicity constraint can be tight and linearly independent of the tight Border constraints \emph{only if} $(i,\tau_{i,j+1}) \succeq (i,\tau_{i,j})$. 
\end{prevproof}



\end{document}